\documentclass[draft,jgrga]{AGUTeX}
\usepackage{graphicx}        % For eps figures, newer & more powerful
\usepackage[usenames,dvipsnames,svgnames]{xcolor}
\usepackage{url}             % For breaking URLs easily trough lines
\usepackage{lineno}
\newcommand{\m}[1]{{{#1}}}
\newcommand{\ssr}{    {\it Space Sci. Rev.}}

\authorrunninghead{JANVIER ET AL.}

\titlerunninghead{COMPARING GENERIC MC AXIS \& SHOCK SHAPES}

\authoraddr{Corresponding author: M. Janvier,
Department of Mathematics, University of Dundee, Dundee DD1 4HN, Scotland,  
       United Kingdom (mjanvier at maths.dundee.ac.uk)}

\begin{document}

\title{{Comparing generic models} for interplanetary shocks and magnetic clouds axis {configurations} at 1~AU}

\authors{M. Janvier,\altaffilmark{1} S. Dasso,\altaffilmark{2,3} P. D\'emoulin,\altaffilmark{4} J.J. Mas\'{i}as-Meza,\altaffilmark{5} N. Lugaz \altaffilmark{6}}

\altaffiltext{1}{Department of Mathematics, University of Dundee, Dundee DD1 4HN, Scotland, 
        United Kingdom}

\altaffiltext{2}{Instituto de Astronom\'\i a y F\'\i sica del Espacio (IAFE), UBA-CONICET, CC. 67, Suc. 28,
       1428 Buenos Aires, Argentina}

\altaffiltext{3}{Departamento de Ciencias de la Atm\'osfera y los Oc\'eanos (DCAO) and Departamento de F\'\i sica (DF), Facultad de Ciencias Exactas y Naturales, Universidad de Buenos Aires, 1428 Buenos Aires, Argentina}

\altaffiltext{4}{Observatoire de Paris, LESIA, UMR 8109 (CNRS), 92195 Meudon Principal Cedex, France}

\altaffiltext{5}{Departamento de F\'isica and IFIBA, Facultad de Ciencias Exactas y Naturales, Universidad de Buenos Aires, 1428 Buenos Aires, Argentina}

\altaffiltext{6}{Space Science Center and Department of Physics, University of New Hampshire, Morse Hall, 8 College Rd, Durham, NH, 03824, USA}

\begin{abstract}
% context 
   Interplanetary Coronal Mass Ejections (ICMEs) are the manifestation of solar transient eruptions, 
which can {significantly} modify the plasma and magnetic conditions in the heliosphere.
{They are often preceded by a shock, and a magnetic flux rope is detected in situ in a third to half of them.} 
The main aim of this study is to obtain the best quantitative shape for the flux rope axis and for the shock surface from {in situ} data obtained during spacecraft {crossings of these structures.}
% methods 
   We first compare the orientation of the flux ropes axes and shock normals \m{obtained from} independent data analyses of the same events, observed in situ at 1~AU from the Sun.  Then, we carry out an original statistical analysis of axes/shock normals by deriving the statistical distributions of their orientations.  We fit the observed distributions using the distributions derived from several {synthetic} models describing these shapes.
% results 
   We show that the distributions of axis/shock orientations are {very sensitive to} their respective shape.
One classical model, used to analyze interplanetary imager data, is incompatible with the in situ data.  Two other models are introduced, for which the results \m{for} axis and shock normals lead to very similar shapes; {the fact that the data for MCs and shocks are independent strengthen this result.} The model which best fit all the data sets has an ellipsoidal shape with {similar aspect ratio values} for all the data sets.
   These derived shapes for the flux rope axis and shock surface have several potential applications. First, these shapes can be used to construct a consistent ICME model. 
   Second, these generic shapes can be used to develop a quantitative model to analyze imager data, as well as constraining the output of numerical simulations of ICMEs. 
   Finally, they will have implications for space weather {forecasting, in particular for} forecasting the time arrival of ICMEs {at the Earth.}

\end{abstract}
%    \keywords{Sun: coronal mass ejections (CMEs), Sun: heliosphere, magnetic fields, Sun: solar-terrestrial relations, Sun: space weather}

\begin{article}
%________________________________________________________________________%%%%%%%%%%%%%%%%%%%%%%%%%%%%%%%%%%%%%%%%%%%%%%%%%%%%%%%%%%%%%%%%%%%%%%%%%%%%%%%%%%%%%
\section{Introduction} %%%%%%%%%%%%%%%%%%%%%%%%%%%%%
\label{sect_Introduction}
  
%{\S\bf ICMEs: I-CME (FR+sheath+shock):} \\
\m{Interplanetary coronal mass ejections (ICMEs) are the interplanetary manifestations of coronal mass ejections (CMEs) and are identified by a number of typical properties that differ from those of the ambient 
solar wind \citep[e.g.,][]{Gosling90,Gosling00,Neugebauer97,Wimmer-Schweingruber06s,Zurbuchen06,Rouillard11b}.
A region of compressed plasma and magnetic field, called the sheath, is 
typically present at the front of ICMEs. When the leading front is faster than the ambient solar wind (above the local fast MHD mode speed), an ICME is preceded by a shock.}

%The interplanetary manifestation of a Coronal Mass Ejection (CME) produces an altered solar wind structure {with different plasma and magnetic field properties \citep[e.g.,][]{Gosling90,Gosling00,Wimmer-Schweingruber06s,Rouillard11b}. When the interplanetary CME (ICME) is faster than the ambient solar wind (above the local fast MHD mode speed), it is preceded by a shock wave \citep[see, e.g., the review by][and references therein]{Gosling90}, through which mass and magnetic field are accreted} from the ambient solar wind, adding magnetised material to its sheath \cite[see e.g., Figure~2 in][]{Zurbuchen06}.

%{\S\bf intro on ICME shape:} \\
An ICME can be distinguished from the ambient solar wind by analyzing
both remote and in situ observations, \m{with remote observations being the privileged method to give hints on their 3D shape.
Nonetheless, coronagraphs and heliospheric imagers still only provide 2D images of the denser parts of CMEs through the Thomson scattering of white-light by free electrons \citep{Howard11,Thernisien11}.
Furthermore, since the location of the scattered light cannot unambiguously be determined along the line of sight, a shape model is typically needed to analyze the observations.}
  
\m{Classical shape models are point-like, or are represented by a sphere surrounding or attached to the Sun \citep[][and references therein]{Lugaz10}.
These models allow the estimation of the propagation direction and the speed of ICMEs, especially when triangulation with different spacecraft is possible \citep{Liu13}.  
More elaborated analytical models have been proposed such as a dense shell located around a flux-rope like shape \citep{Thernisien06}, parametric models of sheath shapes \citep{Tappin09,Howard10} and another parametric shape applied to both sheaths and flux ropes was proposed by \citet{Wood09c}.
% \citet{Thernisien06} proposed a model with a dense shell located on a flux-rope like shape. 
% It was since applied successfully to several events \citep[{\it e.g.}][]{Krall07,Thernisien11b}. 
%\citet{Tappin09} and \citet{Howard10} proposed and applied to observations several analytical models for the sheath shapes.  Finally \citet{Wood09b,Wood09c} proposed other analytical expressions for the sheath and flux rope shapes. 
  All these models were fitted visually to CMEs observed with coronagraphs and/or heliospheric imagers.  However, it remains difficult to quantify the quality of the fit for each of these models, and to confirm whether the derived shape is indeed representative of the CME 3D shape ({\it e.g.} multiple solutions are possible for a given set of CME observations).  Finally, while other techniques have been tested, as reviewed by \citet{Mierla10}, present} 
% Using a combination of numerical simulations and remote observations of coronal structures with white light coronagraph, the denser coronal plasma in front of CMEs appears to be located on a flat croissant-shaped structure \citep[e.g.,][]{Rouillard11b}. 
%However, at larger distances from the Sun, the shape can significantly change due to the interaction of the ICME with the solar wind. 
multi-spacecraft remote observations cannot uniquely determine the 3D shape of the bright front, even when they can be tracked and be unambiguously connected to the CME source \citep{Harrison08,Mostl09,Lugaz11}.

%{\S\bf MCs:} \\
Magnetic Clouds (MCs) can be distinguished inside a fraction of ICMEs, when they are observed in situ. 
A MC is identified as a structure presenting an enhanced magnetic field intensity, a large scale and coherent magnetic field rotation (associated with the passage of a large scale flux rope, FR), 
and low proton temperatures \citep[e.g.,][]{Burlaga81,Dasso05}.
Theoretical models describing the global shape of MCs in the heliosphere have been developed and compared with single spacecraft in situ observations \citep[e.g.,][]{Marubashi07,Hidalgo12}.
However, reconstructing the 3D global FR shape from in situ measurements is an ill-posed problem, as there is no unique solution. Moreover, the models frequently contain many free parameters, so that minimizing the difference between observations and models can sometimes provide several compatible solutions.

%{\S\bf geometry of MCs from multi S/C:} \\
Simultaneous multi-spacecraft observations of a given event can provide a better understanding of the global shape of MCs.
Some studies have shown that the directions of the local axis of a given MC, at well separated locations from different spacecraft, 
are consistent with a smooth global axis of the cloud in the heliosphere (see e.g., \cite{Burlaga81} and Figure 2 in \cite{Ruffenach12}) but this is not always the case (e.g., \cite{Farrugia11}).
However, this kind of simultaneous observations of the same event is not frequent.

%{\S\bf geometry of MCs/ICMEs from numerical simulations:} \\
Numerical simulations have also been used to better understand the propagation and evolution of MCs in the solar wind \citep[e.g.,][]{Riley03,Manchester04b,Lugaz11}.
There are however limitations for emulating some physical processes, such as magnetic reconnection, {the effects of the turbulence on macroscopic structures such as the drag \citep[e.g.,][]{Matthaeus11}}, 
or the dynamical evolution of structures strongly dominated by the magnetic field (such as magnetic clouds). Some ICME propagation properties can be found by combining observations with simulations \citep[e.g.,][]{Lugaz2009,Lugaz11}.

%{\S\bf Our two previous papers on shapes (MCs and shock):} \\
Recently, \citet{Janvier13} analyzed the distribution of MC local orientation from a sample of more than 100 events observed at 1 AU by Wind, and derived the mean shape of the MC axis with an original statistical method. 
A comparable technique {was also used} for deriving the shock surface driven by ICMEs from observations of the shock normal in a sample of more than 250 events observed at 1 AU by the ACE spacecraft \citep{Janvier14b}. 

%{\S\bf What is done in this paper and Roadmap:} \\
In this paper, we apply, extend and combine the techniques previously used \citep{Janvier13,Janvier14b}, with the aim of determining the generic 
shape of the MC axis and the shock surface, combining different samples/databases for MCs and shocks.
In Section~\ref{sect_Observations}, we present the samples of MCs and shocks we use in this statistical study, and provide a comparison 
of the results given for the same events analyzed by different authors. 
We also define the angles used to determine the orientation of the MC axis and normal to the shock surface.
In Section~\ref{sect_Model}, we first set the techniques to derive the axis and shock shapes with the same formalism.
Then, we apply them to various data sets and we compare these results to three analytical models. 
We find that one of the models is the closest to all the observations analyzed, and we test the robustness of our results.
{Then we summarize our main results in Section~\ref{sect_Summary} and outline the implications of our results in Section~\ref{sect_Conclusion}.}

\begin{figure*}  %________________________ FIG ______________________________________    
\centering
\IfFileExists{2columnFigures.txt}{
\includegraphics[width=0.8\textwidth,clip]{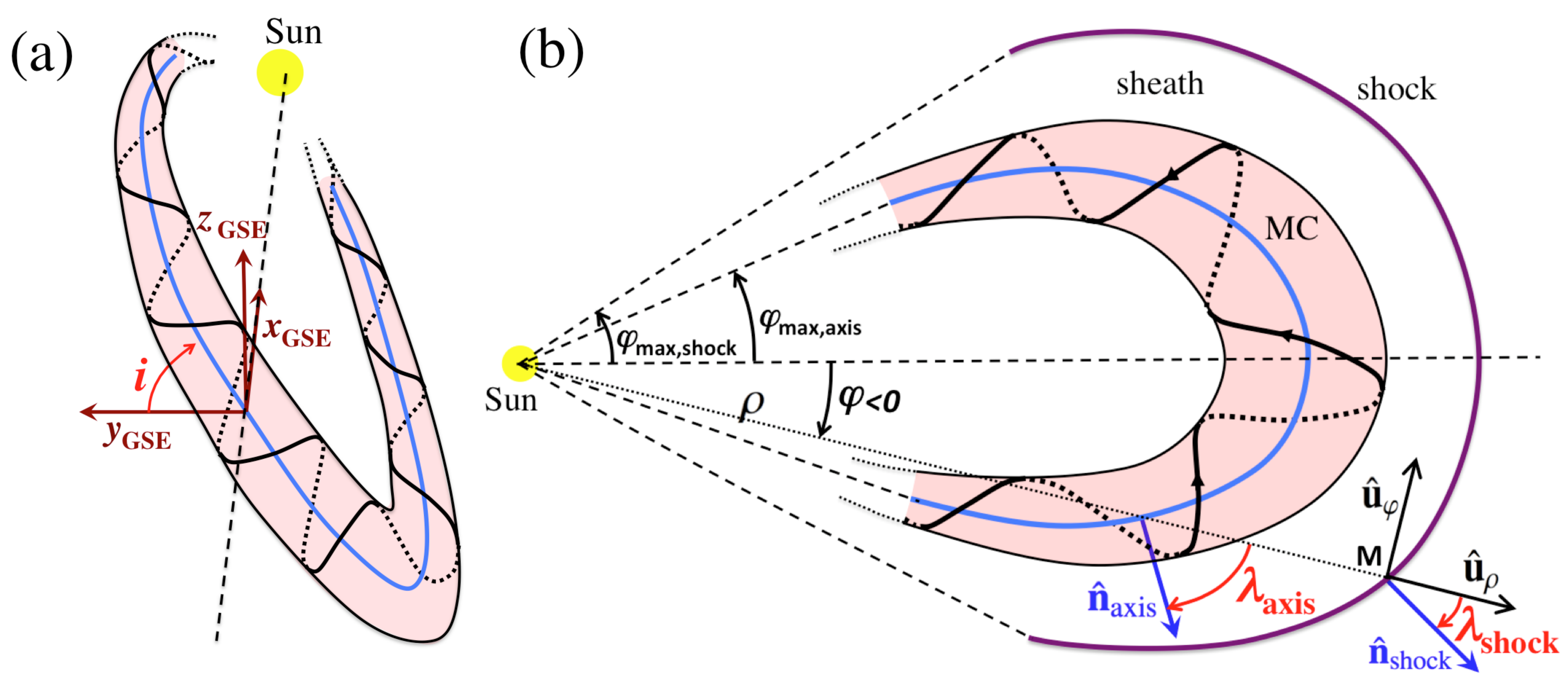}
  } {
\includegraphics[width=\textwidth,clip]{fig_schema_png}
}
\caption{Diagram of a MC and {associated parameters:   
(a) perspective view, (b) side view. The MC axis is drawn in blue, its borders in black and one representative magnetic field line is added. The inclination angle, $i$, is defined by the angle made by the axis direction projected on the $y$-$z$ plane with the $y$ direction (in GSE coordinates).  The maximum angular extension of the MC axis, as viewed from the Sun, is $\varphi_{\rm max, axis}$ while the shock extends up to $\varphi_{\rm max, shock}$. The location angle} $\lambda$ is defined by the angle between the radial direction from the Sun and the normal to the MC axis or the shock. 
}
\label{fig_schema}
\end{figure*}

%%%%%%%%%%%%%%%%%%%%%%%%%%%%%%%%%%%%%%%%%%%%%%%%%%%%%%%%%%%%%%%%%%%%%%%%%%%%%%%%%%%%%
\section{Samples of observed MCs and shocks and definition of the shape parameters} %%%%%%%%%%%%%%%%%%%%%%%%%%%%%
\label{sect_Observations}

\subsection{Sets of observed MCs, similarities and differences}  %%%%%%%%%%%%%%%%%%%%%%%%%%%%%
\label{sect_MC-set}

  %{\S}{\bf --- The MC lists (spacecraft, period, boundaries)} \\
In the present study, we use three lists of MCs where the FR parameters are already determined 
\m{by a common method, namely, the data were fitted with the static, linear force-free cylindrical model, also known as the Lundquist's model \citep[see \textit{e.g.}][]{Lundquist50,Goldstein83,Lepping90}. A detailed description of the model and its limitations can be found in \citet{Lepping90,Lepping03b} and more information on the fitting methods used can be found in \citet{Lynch03} and \citet{Feng07}. }

\citet{Lynch05} analyzed 132 MCs observed nearby Earth by Wind and ACE spacecraft during the period 1995-2003. 
\citet{Feng10} analyzed 62 MCs preceded by a shock and observed by Wind during the period 1995-2007.  
Finally, we use an extended list of events (Table~2 at http://wind.nasa.gov/mfi/mag\_cloud\_S1.html) which is based on the results of \citet{Lepping10} and includes more recent MCs. This list contains the parameters obtained for 121 MCs observed by Wind spacecraft during the period 1995-2009. 
\m{We remove from \citeauthor{Lepping10}'s list the MCs that were crossed too close from their boundaries, limiting the list to 107 MCs \citep[for coherence with previous study, see][]{Janvier13}.}   

  %{\S}{\bf --- The fitting procedure } \\ 
\m{The above fitting method has several limitations.} 
For example, due to its simplicity, the model does not take into account the expansion of MCs, although this could affect the fitted model parameters and the derived quantities (\textit{e.g.} see Tables 3 and 4 in \citet{Nakwacki08}).
Another important limitation is the assumed circular cross \m{section, and as such several non-circular} models have been proposed \citep[\textit{e.g.}][]{Hu02,Vandas05,Hidalgo03}.  
\m{From a statistical study of the closest approach of the spacecraft from the MC axis, using the
\citeauthor{Lepping10}'s list,} \citet{Demoulin13} showed that the MC cross section is flatter in the radial direction (from the Sun) by a third to a half on average. Indeed, some MCs have a relatively \m{flattened} cross section \citep{Vandas05,Antoniadou08,Farrugia11}, while others have a more round cross section, especially at their cores \citep{Hu05b,Liu08,Mostl09,Isavnin11}.
{The effects of a non-circular cross section on the derived properties of MCs (\textit{e.g.} their axis direction) remain to be characterized for large sets of MCs.  So far such lists are only available for the fit with the Lundquist model, so with a circular cross section. } 

  %{\S}{\bf --- Differrences in the fitting procedure } \\
While the three MC {lists used in the present study \citep{Lynch05,Feng10,Lepping10}} are based on the same fitting method to determine the FR parameters, the fit has been applied in different ways to the various MC samples.
For example, \citet{Lynch05} imposed that the spacecraft closest approach to the FR axis is the temporal midpoint of the FR, so that the FR radius is not a parameter of the fit (in contrast with other authors), 
but is instead determined from the FR properties: orientation, mean velocity and impact parameter (defined as the closest distance approach of the spacecraft to the FR axis divided by the FR radius). Among the differences in the method implementation, the difference in the FR boundary selection is, in our opinion, the most important, as it could have large implications on the derived FR orientation \citep[\textit{e.g.} ][]{Dasso06,Ruffenach12}.   The boundaries are typically set where abrupt changes of the magnetic field and plasma parameters are detected \citep[\textit{e.g.} ][]{Dasso07}.   However, {if such changes are not co-located,} then different criteria can provide different boundary locations.
As such, it often occurs that one MC can be defined with different boundary locations depending on the authors \citep[\textit{e.g.}][]{Riley04c,Al-Haddad13}.

An important {physical process during the} evolution of MCs in the solar wind to be considered, and which strongly affects the determination of their boundaries, is their erosion as they travel away from the Sun. As the magnetic field of MCs can reconnect with that of the ambient solar wind, a part of the original FR disappears, so that a {case-dependent fraction of a MC} can have a mixture of both MC and solar wind properties.
This leads to large uncertainties when a FR is actually identified. 
{Thus, a detailed analysis of each MC is needed to determine which remaining region of the FR is} crossed by the spacecraft, and the fit should only be applied to this specific part of the MC.
As an example, {minimum variance and Lundquist's fit can result in significantly} different orientations, without coherence along the axis of a MC observed by four spacecraft \citep[ACE, STEREO A and B, Wind;][]{Farrugia11}, while both methods give consistent results when refined time intervals (\textit{e.g.} considering an eroded FR) are used \citep{Ruffenach12}.  

%Such a detailed analysis was not performed for any of the three readily available lists of MCs used in the present paper, and performing such analysis is outside the scope of this paper.
%Finally, more information on the fitting methods used can be found in \citet{Lepping90, Lepping03b,Lynch03,Feng07}.    
 
\subsection{Sets of observed shocks}  %%%%%%%%%%%%%%%%%%%%%%%%%%%%%
\label{sect_Shocks-set}

   %{\S}{\bf --- Shocks.  data sets} \\
In the present paper, we study the probability distributions of shock orientation parameters from different samples. We use lists of shocks studied in \citet{Feng10} and \citet{Wang10}. The shocks of \citet{Feng10} were observed by Wind spacecraft during the period 1995-2007 and were all located at the front of a MC sheath. The list contains 62 shocks.  
Shock and shock-like events from \citet{Wang10} were observed by ACE spacecraft during the period 1998-2008. They were investigated with the purpose of analyzing their effects on the Earth magnetosphere and ionosphere. In the following, we only retain the well-defined shock events, which represent a total of {216 cases (117 shocks in front of ICMEs and 99 shocks with no detected ICMEs behind).}
 
   %{\S}{\bf --- Shocks.  method to get the normal} \\
Both of these shock studies are based on a shock fitting procedure using the MHD Rankine-Hugoniot relations as developed by \citet{Lin06}. The model used to determine the geometry of the shock is based on the one-fluid anisotropic Rankine-Hugoniot relations. A Monte-Carlo calculation and a minimization technique between the generated models and observations were also similarly used.  However, while the basic technique is the same in both studies, there are variations affecting the computed shock parameters. Firstly, the data are from two different spacecraft, involving different measurements with their own specificities. Secondly, the results depend on the selected time intervals, both for the upstream and downstream regions of the shocks, which are used to select the magnetic field and plasma parameters used in the minimisation technique. Finally, the results also depend on the terms included in the Rankine-Hugoniot relations \citep[see the two methods studied in][]{Lin06}.  As a consequence, differences between the results of both studies have to be expected when the same shock is analyzed. We analyze these differences in Section~\ref{sect_Precision}.

\subsection{Angles defining the MC axis and shock normal}  %%%%%%%%%%%%%%%%%%%%%%%%%%%%%
\label{sect_Angles_Normal}

  %{\S\bf GSE. Latitude and longitude} \\  
The directions of the FR local axis and the shock normal 
are obtained from the fitting and minimisation procedures for both MCs and shocks, respectively.  For in situ observations at 1~AU, \textit{i.e.} nearby Earth, unit vectors used to express directions are classically defined in the GSE system of reference. Setting a spherical coordinate-system with a South-North polar axis, the orientation vector is classically defined by its latitude and longitude. 
However, the polar axis direction is singular, as it corresponds to any values of longitude. Therefore, as the absolute value of the latitude becomes larger, any small change in the vector orientation leads to a large uncertainty in the longitude determination. {This is a problem when characterizing the orientation of the axis of a MC, which may be highly inclined to the ecliptic. However, MC axes are found to rarely lie in the radial direction. Thus,} \citet{Janvier13} introduced a new spherical coordinate-system, with a polar axis along the Sun-Earth direction (along $-\hat{\bf{x}}_{GSE}$).  Projecting the unit vector on a plane perpendicular to $\hat{\bf{x}}_{GSE}$, they introduced the inclination angle $i$ on the ecliptic. 
It is measured from the West-East direction ($\hat{\bf{y}}_{GSE}$) in a clockwise direction (when looking toward the Sun, {see Figure~\ref{fig_schema}a}).
A second angle was defined as the location angle $\lambda$, which measures the angle between the FR axis direction and the ortho-radial.   Equivalently, $\lambda$ is also the angle between the radial direction and the normal to the FR axis ($\hat{\bf{n}}_{\rm axis}$ located within the plane $i=$ constant, {Figure~\ref{fig_schema}b}). 
Then, for both cases (shock normal and MC axis), $\lambda$ is defined in the same way, as the angle between -$\hat{\bf{x}}_{GSE}$ and $\hat{\bf{n}}$ ($\hat{\bf{n}}_{\rm shock}$ or $\hat{\bf{n}}_{\rm axis}$) and both $\lambda_{\rm axis}$ and $\lambda_{\rm shock}$ measure the location of the spacecraft crossing if the {MC/shock} axis shape is known. 
For $\lambda_{\rm axis}$ and $\lambda_{\rm shock}$ values close to zero, the spacecraft crossing is close to the apex of each structure, while both  $\lambda_{\rm axis}$ and $\lambda_{\rm shock}$ increase as the crossing is more on the {flank} of the encountered structure, therefore defining $\lambda$ as a ``location'' angle (see Section~\ref{sect_Model}).

  %{\S\bf $i$ and $\lambda$} computations\\  
In summary, ($\lambda, i$) defines a new spherical coordinate system where $-\hat{\bf{x}}_{GSE}$ is the polar axis. Both angles $\lambda$ and $i$ are related with the latitude $\theta$ and the longitude $\phi$ of $\hat{\bf{n}}_{\rm axis}$ or $\hat{\bf{n}}_{\rm shock}$ by:  
\begin{linenomath}
\begin{eqnarray}
  \sin \lambda &=& -\cos \phi ~\cos \theta    \label{eq_lA} \,,\\
  \tan i &=& \tan \theta ~/~ |\sin \phi |  \label{eq_iA} \,.
  \end{eqnarray} 
  \end{linenomath}
{Then, $\lambda$ and $i$ are both functions of $\phi$ and $\theta$.
However, with MC data \citep{Janvier14c} as well as for shocks (not shown), 
$\lambda$ is mostly correlated with $\phi$, and $i$ with $\theta$.}

\subsection{Precision of the axis and shock normal directions}  %%%%%%%%%%%%%
\label{sect_Precision}

   %{\S\bf Test: Associations}\\
The typical level of precision that can be determined for $\hat{\bf{n}}_{\rm axis}$ and $\hat{\bf{n}}_{\rm shock}$ is investigated in more detail below, by comparing the orientation parameters for events that are common to different data sets.
To do so, we associated events by comparing the times defining the boundaries of MCs or shocks. 

The time window for the association of MCs from different lists was set to 10~hours, for both front and rear boundaries. We chose a rather large window, as the definition of the MC boundaries is dependent on the plasma and magnetic data used, as well as on the authors' selection criteria. However, we also computed an association with a smaller time window (\textit{e.g.} 5 hours), which provided comparable results but with less cases. Extending the time window to more than 10 hours, however, leads to {a few MCs of one list associated to two MCs of another list}, and therefore wrong results.

For shocks, the time window was set to 1~hour, as shocks are better temporally localized than MC boundaries. We also included in this time window the delay between ACE and Wind observations. Note that further extending this time window (\textit{e.g.} 2 hours) only provides a few more associations.

\begin{figure*}  %________________________ FIG ______________________________________    
\centering
\IfFileExists{2columnFigures.txt}{
\includegraphics[width=0.48\textwidth,clip]{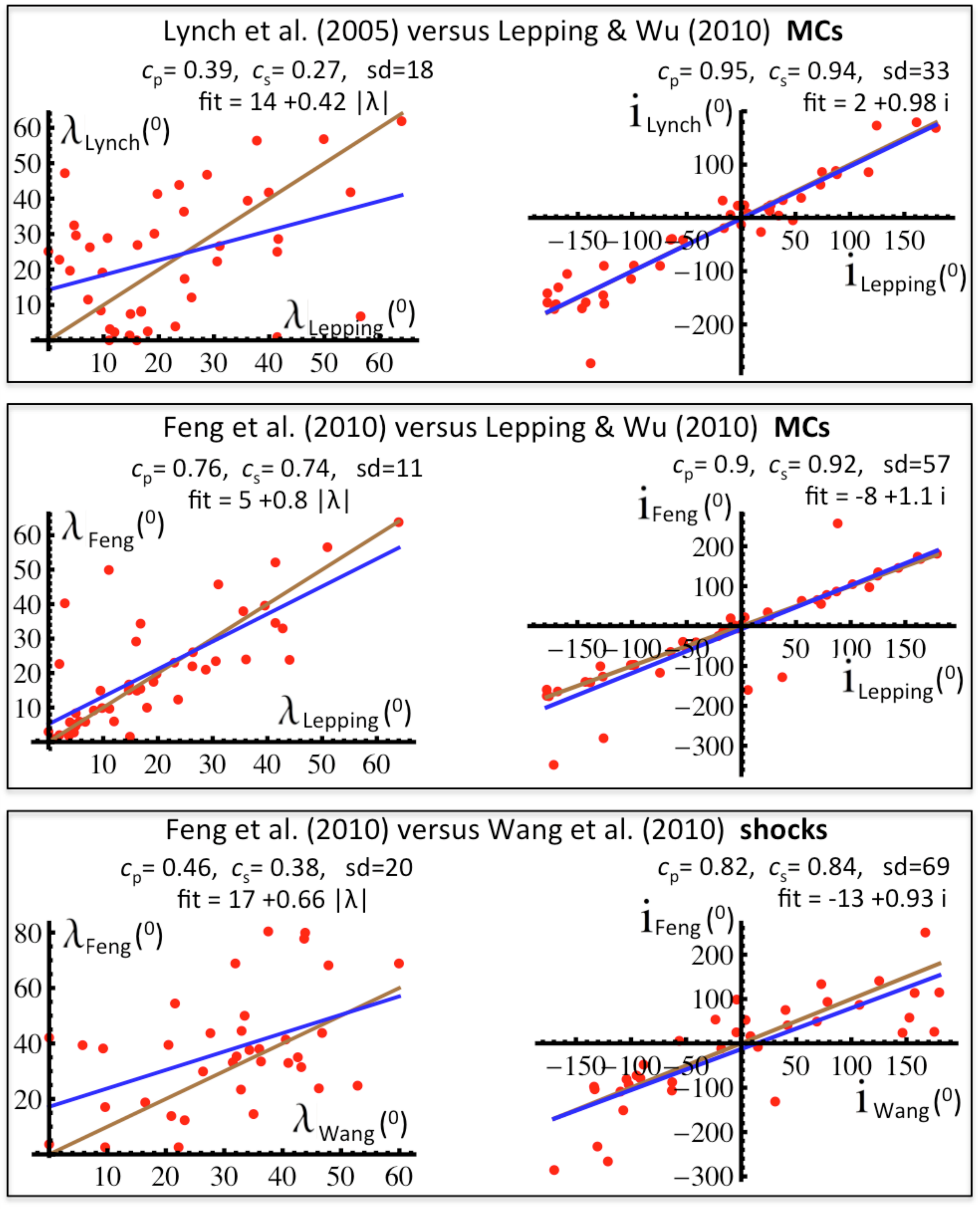}
  } {
\includegraphics[width=0.6\textwidth,clip]{fig_corel_lambda_i_png}
}
\caption{Correlations between the \m{absolute value of the} location angle $\lambda$ (defined in Figure~\ref{fig_schema}{b}) values (left panels), \m{and correlations between the} inclination angle values, $i$ (right panels) obtained for the same cases in different samples of MCs and shocks. 
\m{We show the values of $|\lambda|$ (noted $\lambda$ for conveniency) for the MCs for a better comparison with the shocks (for which the sign of $\lambda$ cannot be determined).} The two first rows are the correlations for MCs, with the orientation parameters found by Lepping and Wu (2010) reported on \m{the horizontal axis.} The values found by Lynch \textit{et al.}  (2005, top) and Feng \textit{et al.} (2010, middle) are reported on \m{the vertical axis.}
The last row represents the correlation between the parameters obtained by associating the shocks studied by Feng \textit{et al.} (2010) with those by Wang \textit{et al.} (2010).  
The parameters $c_{\rm p}$ and $c_{\rm s}$ are the Pearson and Spearman ranking-correlation coefficients, and "sd" is the standard deviation, in degree, between the two data sets (red points).  
The blue line is the least square fit for the data, while the brown line represents 
%\m{equal values of $\lambda $ or $i$ for each set of events. 
\m{{the identity function (i.e., equal values of $\lambda $, or $i$, for each set of events).} 
These results show a correlation both for $\lambda$ and $i$ between different data sets, but with a large dispersion as indicated by "sd".}
}
\label{fig_correlation}
\end{figure*}

   %{\S\bf \% of common cases}\\
The Lepping {and Wu}'s list has 45\% and 49\% {of MCs in common with the Lynch's and Feng's list, respectively}, 
and the Wang's list has only 14\% of shocks in common with the Feng's list {(which is limited to shocks associated to MCs)}. 
This indicates that we can compare the results obtained for the same events, while also having a significant number of independent events.

   %{\S\bf Compare correlations of $\lambda$ and $i$}\\
The relations between $\lambda$ and $i$ for the same events in different pairs of event lists are analyzed in Figure~\ref{fig_correlation}.  \m{We show there the values of $|\lambda|$, \textit{i.e.} we have grouped together the values of $\lambda$ obtained in the west and east legs of the MCs, increasing the statistics (this is also justified by a weak asymmetry between the two MC legs, as found in \citealt{Janvier13}). Moreover, this allows the direct comparison of the results obtained for the MCs with those obtained for the shocks, as the sign of $\lambda$ cannot be determined for the latter. For conveniency, we mark in the following and in all graphs $\lambda$ instead of $|\lambda|$.}
The data are fitted by a linear function (in blue), which provides an easy visual interpretation {of} how close two results are for one same event. 
We also report two correlation-ranking coefficients, the Pearson ($c_{\rm p}$) and Spearman ($c_{\rm s}$) ones, as well as the standard deviation (sd) between the two data sets. As a whole, the location angle $\lambda$ (left panels) between two samples is not as well correlated as the inclination angle $i$ (right panels).  $\lambda$ also has a larger global bias since the linear fit (blue line) is located further away from the {identity} line (in brown) than for the $i$ angle, for which both lines almost coincide. {Finally,} $\lambda$ has a lower standard deviation than $i$ {by a factor 2 to 5}, 
but this is not a striking result considering that the {range of variation for $i$ ($\approx [-150^\circ, 150^\circ]$) is five times larger than the range of variation for $\lambda$ ($\approx [0^\circ, 60^\circ]$)}
{so that the standard deviation relative to the range of variation is broadly comparable for $\lambda$ and $i$}.

   %{\S\bf Correlations of $\lambda$: more }\\
For MCs, the quality of the fit is measured by $\chi$, the square root of the chi-square function between the measurement of the magnetic field inside the FR and the fitted profile.  The values of both $\chi$ and the impact parameter, $p$, could a priori affect the precision of $\hat{\bf{n}}_{\rm axis}$.  In fact, we found no significant improvement in the correlation-ranking coefficients, nor in the standard deviation values, when we analyzed sets of data containing only lower $\chi$ values. 
Next, by taking the common cases in Lepping {and Wu}'s and Feng's lists of MCs, we found that both the correlation coefficients and the standard deviation value are not significantly affected by $|p|$.  However, if Lynch's MCs are associated with either Lepping {and Wu}'s or Feng's MCs, the correlation coefficients significantly increase as the maximum value of $|p|$ decreases from $1$ to $0.5$. 
{Thus,} the results of \citet{Lynch05} depend more on the magnitude of the estimated $|p|$ values than for the other two lists.  

   %{\S\bf Interpretation on the dispersion}\\
The large dispersion in the values of $\lambda$ found by different authors, especially for its lower values, is not straightforward to interpret.
Indeed, for low values of $\lambda$ (\textit{i.e.} near the apex), one would expect that shocks, and similarly MCs, have a {better determined normal as the spacecraft crosses its structure orthogonally}.
These expectations appear to be not true considering the results found in Figure~\ref{fig_correlation}. A possible candidate to {account for such an effect may be the determination of the flux rope boundaries. Thus,} the location of the MC/shock crossing (as quantified by $\lambda$) is not an important parameter in determining the precision of the normal/axis direction.

   %{\S\bf Conclusion }\\
In conclusion, the large dispersion of $\lambda$ and $i$ angles between different MC lists is consistent with the results of \citet{Al-Haddad13} who tested the results of a larger variety of methods but on a more limited sample of MCs.  This large dispersion, both for the MC axis and shock normal directions, does not allow the comparison of these directions for individual MCs \citep[as attempted by][]{Feng10}. 
However, {we will show in Section~\ref{sect_Model} that the distributions of $\lambda$ provide robust information} on the MC axis and the shock shapes.

\subsection{Probability distributions of the orientation parameters}  %%%%%%%%%%%%%%%%%%%%%%%%%%%%%
\label{sect_Probability}

\begin{figure*}  %________________________ FIG ______________________________________    
\centering
\includegraphics[width=\textwidth,clip]{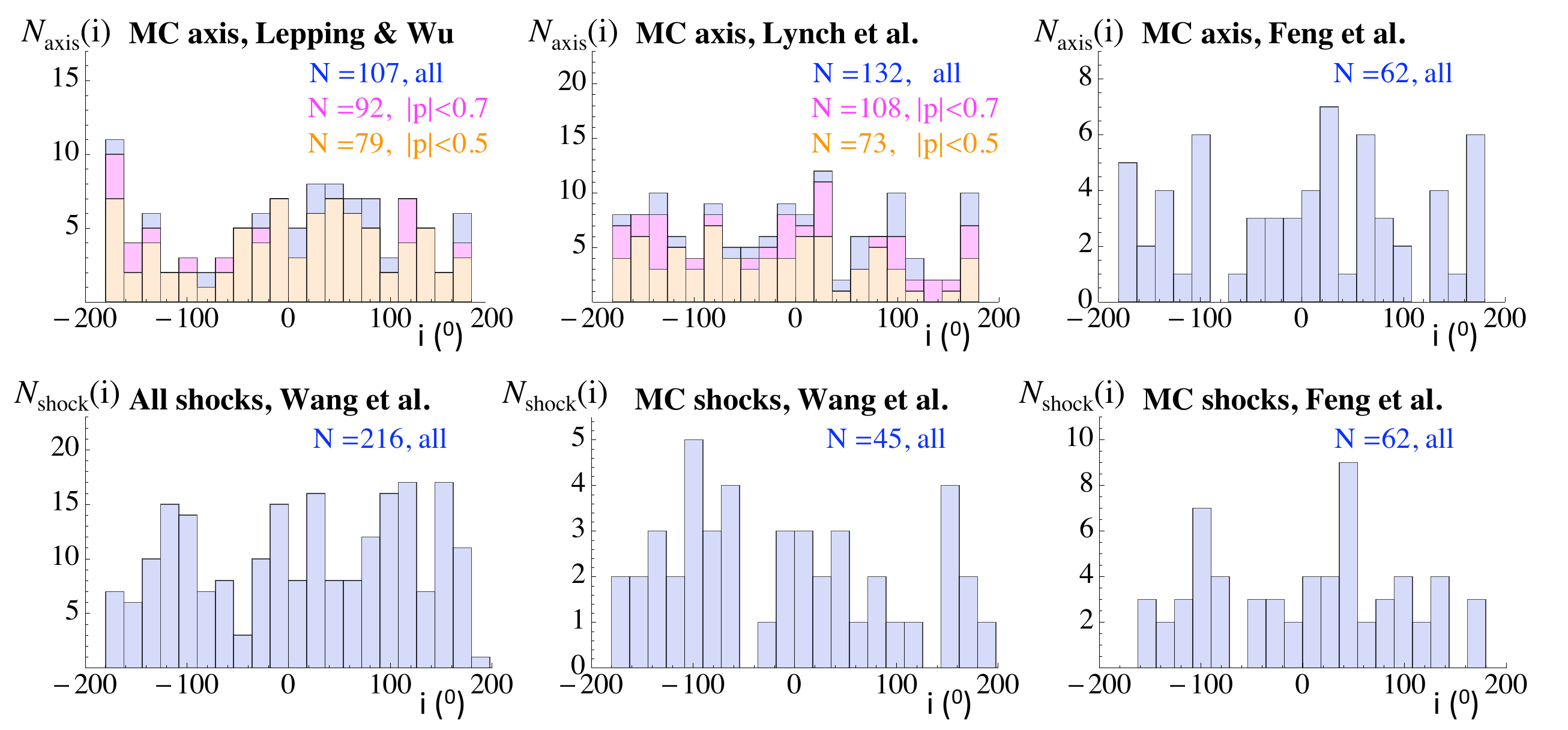}
\caption{Histograms of the inclination angle $i$ (defined in Figure~\ref{fig_schema}a) %, in degree) 
for different data sets as written in the top labels. $N$ is the total number of cases in the histogram. $N_{\rm axis}$ ($N_{\rm shock}$) is the number of MCs (shocks) in each bin, plotted versus $i$ for the MC axis (top row) and the shock normal (bottom row). The 20 bins are regularly spaced between $i =-180$ and $180 \ensuremath{^\circ}$. 
The top left and middle panels show the total number of cases in blue, while the overplotted histograms show cases constrained to an impact parameter $|p|$ lower than a given threshold ($|p|<0.7$ in magenta and $|p|<0.5$ in orange). 
This way, the bars in blue are always the longuest ones in the histogram, as they correspond to the number of cases without constraints on $|p|$, and are covered with the magenta and light orange bars.
The bottom left and middle panels show the distribution of $i$ for the shocks reported in \citet{Wang10}, either taken all together (left) or with the 
\m{constraint that a MC was observed following the shock (middle). These histograms all show no global tendency with $i$ (within the statistical fluctuation limit).} 
 }
\label{fig_prob_i}
\end{figure*}

\begin{figure*}  %________________________ FIG ______________________________________    
\centering
\includegraphics[width=\textwidth,clip]{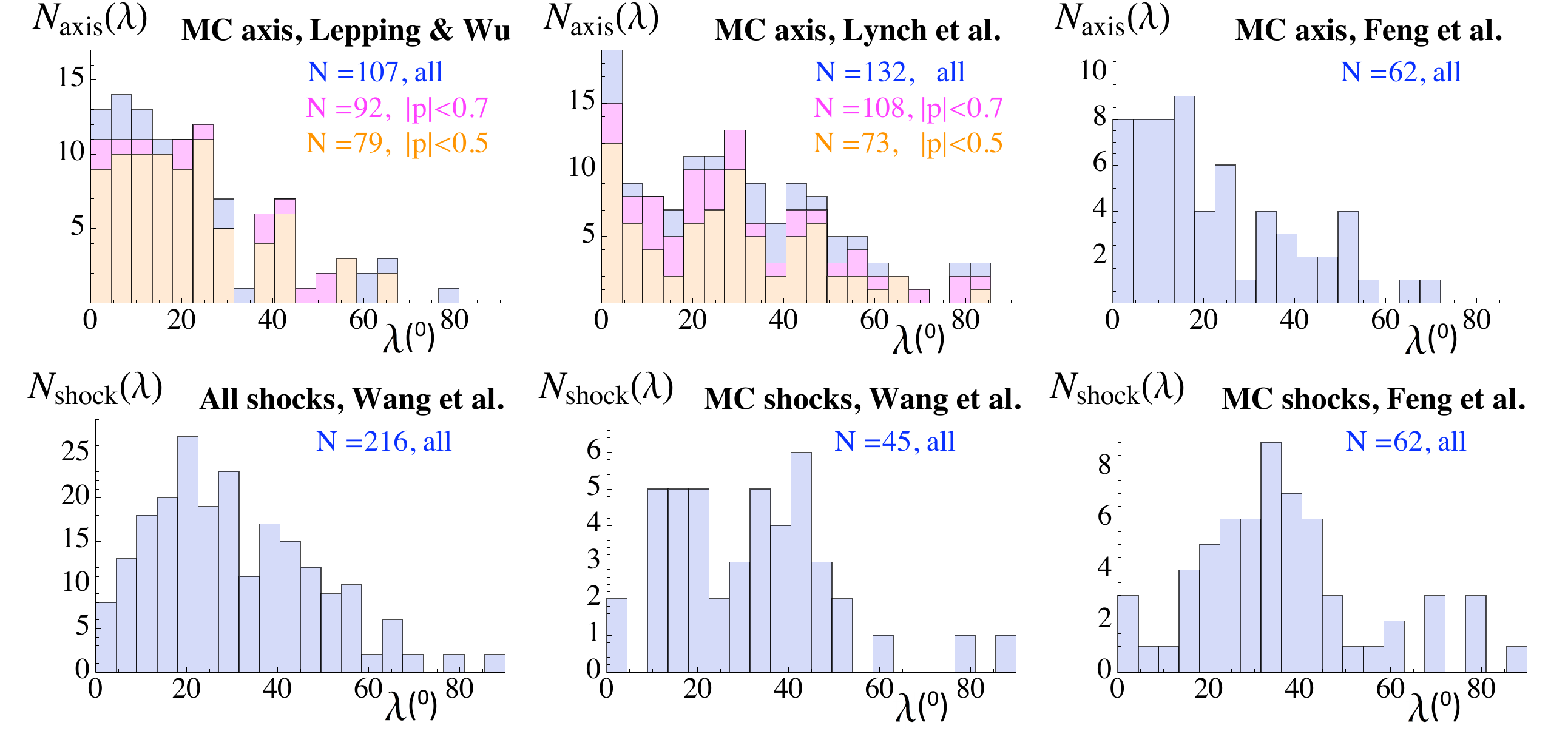}
\caption{Histograms of the location angle $\lambda$ (defined in Figure~\ref{fig_schema}b) %, in degree) 
for different data sets as written in the top labels. $N$ is the total number of cases {in the histogram}. $N_{\rm axis}$ ($N_{\rm shock}$) is the number of MCs (shocks) in each bin, plotted versus $\lambda$ for the MC axis (top row) and the shock normal (bottom row). The 20 bins are regularly spaced between $\lambda =0$ and $90 \ensuremath{^\circ}$. 
\m{The color convention is the same as in Figure~\ref{fig_prob_i}. These histograms are examples of distributions of $\lambda$ used to constrain the MC axis and shock shapes. } 
}
\label{fig_prob_lambda}
\end{figure*}

   %{\S\bf Probability distribution of $i$ }\\
Analyses of the inclination angle~$i$ for the 107 MCs detected by Wind \citep{Lepping10} and for the {216} shocks detected by ACE \citep{Wang10} were performed by \citet[][respectively]{Janvier13,Janvier14b}.  
Taking into account the statistical fluctuations of limited samples, no global tendency was found for $i$ (its probability distribution appeared to be uniform for both MC and shocks). 
We confirm this isotropy of inclination in the present paper by analyzing the histograms of $i$ for the MCs analyzed by \citet{Lynch05} and \citet{Feng10} and for the shocks analyzed by \citet{Feng10}, 
{as shown in Figure~\ref{fig_prob_i} where no global tendency is present over the statistical fluctuations \m{both for the full sets (blue histograms) as well as with restricted sets defined with a restricted range of the impact parameter (pink and light brown histograms) as described at the end of this section}.  
We selected 20 bins in the histograms as a compromise between statistical fluctuations in each bin and a visualization of the variations with $i$.}

   %{\S\bf Interpretation of $\pobs (i)$ }\\
We interpret this isotropy of inclination as follows. The MCs and the majority of the shocks observed at 1~AU are generated by CMEs launched from the Sun. 
\m{During a time period of several years, as covered by the data sets, the Sun launched CMEs from a large number of unrelated source regions with diverse orientations. 
If there is no preferential orientation in the sources, a spacecraft is expected to cross the MCs and shocks associated with the CMEs with an approximative uniform probability of $i$.  As such, the above results 
%\m{are compatible with nearly uniform distributions of $i$. This} 
indicate that there is no privileged direction of $\hat{\bf{n}}_{\rm axis}$ and $\hat{\bf{n}}_{\rm shock}$ around the Sun-apex line (within the limits of the statistical fluctuations). }

   %{\S\bf Probability distribution of $\lambda$ }\\
In contrast, the probability distributions of $\lambda$ (Figure~\ref{fig_prob_lambda}), for both MCs and shocks, are largely non-uniform, even when considering statistical fluctuations (which are typically of the order of $\sqrt{n_c}$ where $n_c$ is the number of cases in an histogram bin).
% We select 20 bins as a compromise between averaging the statistical fluctuations and resolving the distribution dependence with $\lambda$.  
All the MC axis distributions are globally decreasing functions of $\lambda$ (except for the \m{narrow peak at $\lambda \approx 30 \ensuremath{^\circ}$ for Lynch \textit{et al.}, however this peak may not be statistically significant). Then the axis distributions are different than the shock distributions, which  all} peak around $\lambda \approx 30\ensuremath{^\circ}$ (Figure~\ref{fig_prob_lambda}).  Even though subgroups \m{of events} lead to larger statistical fluctuations, the shape of the distributions remains almost the same.  For example, similar histograms are found when considering subgroups of MCs with different fitting qualities \citep[see \textit{e.g.} Figure~5 in][]{Janvier13}, or when considering subgroups of shocks corresponding to different categories of ICMEs \citep[see \textit{e.g.} Figure~5 in][and Figure~\ref{fig_prob_lambda} \m{for shocks associated to MCs}]{Janvier14b}.

   %{\S\bf Effect of impact parameter }\\
The quality of the estimation of the FR axis direction decreases with the absolute value of the impact parameter $|p|$ \citep[\textit{e.g.} ][]{Riley04c}, which is related to the distance of the spacecraft trajectory to the FR axis.  We investigated the effect of $|p|$ on the above distributions, and found no significant effect for both parameters $i$ and $\lambda$: {the histograms with $|p|<0.7$ (pink) and $|p|<0.5$ (light brown) have a similar dependence with $i$ and $\lambda$ than when all cases are considered (blue) as shown in Figures \ref{fig_prob_i} and \ref{fig_prob_lambda}.} This is because large $|p|$ values are present in all the values taken by $i$ and $\lambda$, so that the general tendencies remain the same. The sample that has the largest relative number of cases with $|p|>0.5$ is that of \citet{Lynch05}, with 45\% of the cases (Figure~\ref{fig_prob_lambda}, {top middle panel}).  We anticipate that the conclusion derived from the general properties of the orientation parameters distributions will be weakly affected by the (non-)inclusion of the cases with large $|p|$ values.

%%%%%%%%%%%%%%%%%%%%%%%%%%%%%%%%%%%%%%%%%%%%%%%%%%%%%%%%%%%%%%%%%%%%%%%%%%%%%%%%%%%%%
%\section{Synthetic models for typical axis and shock shapes} %%%%%%%%%%%%%%%%%%%%%%%%%%%%%
\section{Best synthetic models to fit in situ data} %%%%%%%%%%%%%%%%%%%%%%%%%%%%%
\label{sect_Model}

  In this section, we explore the properties of the shapes given by different synthetic models, for both the axis of MCs and the shock {fronts.} These properties are directly compared with those given by in situ data, allowing us to select the best models to reproduce the mean shape of both MC axis and shocks.
To do so, the probability distributions of the parameters describing these synthetic models are least-square fitted to the distribution of the observed MC axis and shock normal. 
 
\subsection{Generic equations for axisymmetric models} %%%%%%%%%%%%%%%%%%%%%%%%%%%%%
\label{sect_spherical-geometry}

  %{\S\bf --- Geometry, $\rho (\varphi)$} \\
We first start with the description of the basic and general equations that we use to describe the models. These equations were previously introduced for the MC axis and shock front \citep{Janvier13, Janvier14b} using spherical coordinates ($\rho,\Phi,\varphi$) centered on the Sun (S). For a given point on the shock surface, $\Phi$ is the angle defining the position around the Sun-apex line, while for the MC axis, $\Phi$ is constant since the axis is supposed to be contained in a plane for a given event.
 %this was not correct since this figure is not showing the same $\Phi$:
 % {(see left panel of Figure 1 in \citealt{Janvier13})}.

For the shock front, we suppose in the present work a symmetry of rotation, so that the shock surfaces are all independent of $\Phi$. This simplifies the expression for $\rho$, the distance between the Sun (S) and the spacecraft crossing point (M). Then, the vector from the Sun to the point M can be written as a function of $\varphi$ as (Figure~\ref{fig_schema}b):  
  \begin{linenomath}
  \begin{equation} \label{eq_OM}
  \vec{SM} = \rho (\varphi) ~\hat{\bf u}_{\rho}  \, .
  \end{equation} 
  \end{linenomath}

  %{\S\bf --- Compute $\lambda$} \\
The location angle $\lambda$ is defined as the angle between the radial direction from the Sun (along $\hat{\bf u}_{\rho}$) and the local normal ({$\hat{\bf{n}}$}) to the axis/shock shape. 
$\lambda$ is related to $\rho (\varphi)$ as:   
  \begin{linenomath}
  \begin{equation} \label{eq_tanLambda}
  \tan \lambda = \frac{\hat{\bf{n}}\cdot\hat{\bf u}_{\varphi}}{\hat{\bf{n}}\cdot\hat{\bf u}_{\rho}} 
           = -\frac{\rm{d} \ln \rho}{\rm{d} \varphi}   \, .
  \end{equation}
  \end{linenomath}
The above two expressions remain the same for both the shock and the MC axis.

\begin{figure*}  %________________________ FIG ______________________________________    
\centering
\includegraphics[width=\textwidth,clip]{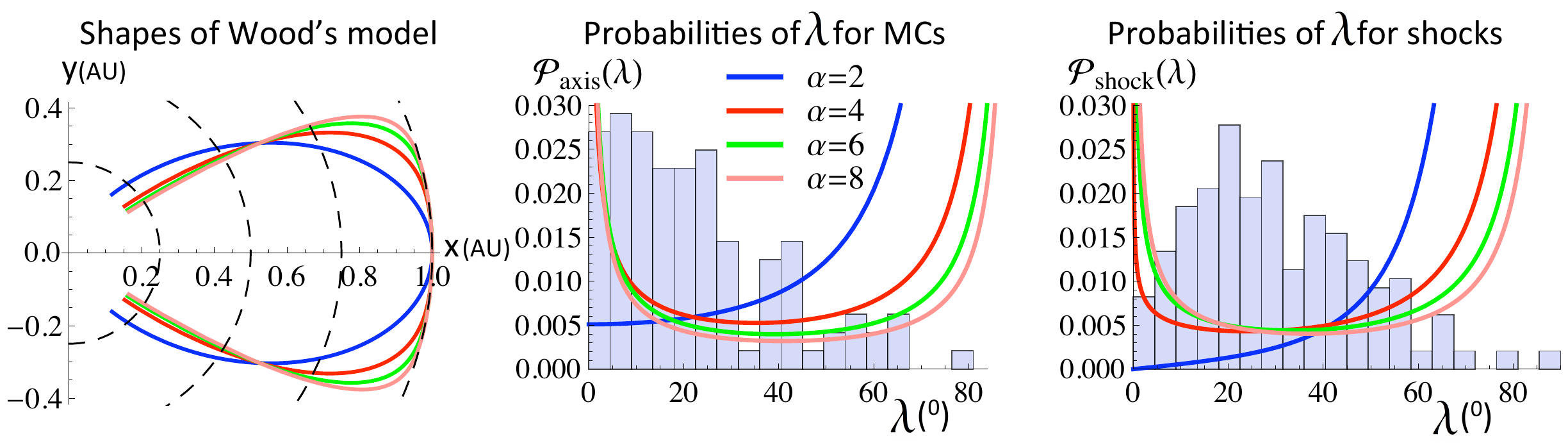}
\caption{Shapes and probabilities per unit $\lambda$, in degree, for the Wood's model (Section~\ref{sect_Mod_Wood}). 
The left panel shows the MC axis or the cross-section of the shock in a plane including the Sun {(located at the origin)} and the axis/shock apex, as defined by Equation (\ref{eq_rho_Wood}) and limited to $\rho >\rho_{\rm min} =0.2$.  
%The X and Y coordinates 
The colored lines are for four different values of the $\alpha$ parameter.  The corresponding probabilities of $\lambda$ are shown in the middle and right panels together with, in the background, the observed probability drawn with histograms for MCs analyzed by Lepping \& Wu (middle panel) and for shocks analyzed by Wang et al. (right panel). \m{The large differences between the colored lines representing Wood's model and the histograms from the observed distribution show that the model is not consistent with the observations.}
}
\label{fig_wood}
\end{figure*}

  %{\S\bf --- Extension} \\
  The shock surface extends from $\varphi=0$ at the apex to $\pm \varphi_{\rm max, shock}$ on the {flank}, with a symmetry of rotation around the Sun-apex line (Figure~\ref{fig_schema}{b}).  
  For consistency, the axis shape is also symmetric around the apex and it extends up to $\varphi=\pm \varphi_{\rm max, axis}$.
 In both cases, $\lambda$ is included in the interval $[0,90 \ensuremath{^\circ} ]$, with $\lambda=0$ at the apex and $\lambda$ approaching $90 \ensuremath{^\circ} $ in the legs of the MC axis/{flank} of the shock {front}.

  %{\S\bf --- Compute $\Pl$} \\
Supposing that $\rho (\varphi )$ {is a} decreasing function of $\varphi$ (so that the shape $\rho (\varphi)$ is concave towards the Sun, in agreement with observations and simulations {\citep[e.g.,][]{Cane85,Lugaz14}}), Equation(\ref{eq_tanLambda}) implies that $\lambda$ is a monotonously increasing function of $\varphi$.  It implies that spacecraft crossings in the range $\varphi \pm \rm{d} \varphi$ correspond to the unique range $\lambda \pm \rm{d} \lambda$.  The conservation of the number of cases implies
\begin{linenomath}
  \begin{equation} \label{eq_conservation}
  \mathcal{P}_{\varphi} (\varphi) ~\rm{d} \varphi = \mathcal{P}(\lambda) ~\rm{d} \lambda  \, ,
  \end{equation}
\end{linenomath}
where $\mathcal{P}_{\varphi} (\varphi) ~\rm{d} \varphi$ is the probability of spacecraft crossings in the interval {$\varphi \pm \rm{d} \varphi/2$ and $\mathcal{P}(\lambda) ~\rm{d} \lambda$ the probability in the interval $\lambda \pm \rm{d} \lambda/2$. } 
  
  %{\S}{\bf --- Hypothesis on the uniform crossing. Define $\mathcal{P}_{\varphi} (\varphi)$} \\
A spacecraft located at 1~AU, over several years, will cross a large number of ICMEs launched from the Sun from a broad range of latitudes and longitudes with no privileged direction {of propagation}.  It implies that one can suppose a nearly uniform distribution of crossings.  Then, the MC axis is expected to be detected with a uniform distribution of $\varphi$.  The normalization of the total probability to unity implies $\mathcal{P}_{\varphi} (\varphi)=1/ \varphi_{\rm max}$. This probability is different for shocks however, since they extend as a surface: the probability of detection in the range $\varphi \pm \rm{d} \varphi$ is proportional to the corresponding fraction of the cross section of the sphere of radius 
$D=1$~AU centered on the Sun, so $\mathcal{P}_{\varphi} ~\rm{d} \varphi \propto 2 \pi D^2 \sin \varphi \, \rm{d} \varphi $.  The coefficient of proportionality is again found by normalizing the total probability to unity.
Summarizing, the probability of crossing axis/shock is:
\begin{linenomath}
  \begin{eqnarray} 
  \mathcal{P}_{\varphi} (\varphi) 
  &=& 1/ \varphi_{\rm max}  \hspace{0.065\textwidth} \mbox{\rm for MC axis} 
          \label{eq_pphi_axis} \, , \\
  &=& \frac{\sin \varphi}{1-\cos \varphi_{\rm max}} \quad \mbox{\rm for shocks} 
          \label{eq_pphi_shock} \, .
  \end{eqnarray}
\end{linenomath}
  %{\S}{\bf --- Compute $\mathcal{P}(\lambda)$} \\
With the above result for $\mathcal{P}_{\varphi} (\varphi)$, Equation~(\ref{eq_conservation}) defines $\mathcal{P}(\lambda)$ when
$\rm{d} \varphi/\rm{d} \lambda$ is known, \textit{i.e.} when the shape is known since the derivation of Equation~(\ref{eq_tanLambda}) with respect to $\lambda$ defines $\rm{d} \varphi/\rm{d} \lambda$.
All in all, $\mathcal{P}(\lambda)$ writes:
\begin{linenomath}
  \begin{equation} \label{eq_Pl}
  \mathcal{P}(\lambda) = \mathcal{P}_{\varphi} (\varphi) 
        \frac{1}{\cos^2 \lambda ~({-}\rm{d}^2 \ln \rho / \rm{d} \varphi^2)} \, .
  \end{equation}
  \end{linenomath}
  This expression is common for both the MC axis and the shock, as $\mathcal{P}_{\varphi} (\varphi)$ can then be replaced by its proper expression from Equation~(\ref{eq_pphi_axis}) or Equation~(\ref{eq_pphi_shock}).
Since $\varphi$ can be expressed as a function of $\lambda$ with Equation~(\ref{eq_tanLambda}) when $\rho (\varphi)$ is specified, $\mathcal{P}(\lambda)$ in Equation~(\ref{eq_Pl}) can be written as a function of $\lambda$.

  %{\S}{\bf --- Compare models to observations} \\
A quantitative comparison between two probability distribution functions, with one corresponding to the real shape $\mathcal{P}_{\rm real}$ (\textit{i.e.} estimated from observations) and the other corresponding to a given synthetic model $\mathcal{P}_{\rm model}$, can be done by computing the distance between the two probability curves, as:
\begin{linenomath}
  \begin{equation} \label{eq_dist_two_curves}
  \rm dist(\rm real, \rm mod.) = \sqrt{ \frac{1}{90} \int_{0}^{90} 
                 \Big(\mathcal{P}_{\rm real} (\lambda)-\mathcal{P}_{\rm model}(\lambda) \Big)^2  \rm d\lambda        } \, .  
  \end{equation}
\end{linenomath}
{With a given analytical model of $\rho(\varphi)$ for the MC/shock shape, Equation~(\ref{eq_Pl}) provides a continuous function of $\lambda$.} 
However, since the observed probabilities are binned, the comparison requires to also bin the probability obtained from the synthetic model (the probability function of this binned model is then noted $\mathcal{P}_{\rm bmod} (\lambda )$).   
Then, Equation~(\ref{eq_dist_two_curves}) is transformed into:
\begin{linenomath}
  \begin{equation} \label{eq_diff}
  \rm{diff} (obs., mod.) = \sqrt{ \frac{1}{n_{\rm b}} \sum_{i=1}^{n_{\rm b}} 
                 \Big(\mathcal{P}_{\rm obs} (\lambda_i)-\mathcal{P}_{\rm bmod}(\lambda_i) \Big)^2                 } \, ,
  \end{equation}
  \end{linenomath}
where $n_{\rm b}$ is the number of bins of the observed probability $\mathcal{P}_{\rm obs}$.

\subsection{Wood's model} %%%%%%%%
\label{sect_Mod_Wood}

  %{\S}{\bf --- Definee $\rho_w (\varphi)$} \\
We first start our investigation of the most appropriate synthetic model with the Wood's model. In a series of papers, Wood and co-authors used images from both STEREO spacecraft to visually fit the ICME front and/or the leading and trailing edges of flux ropes with the following model:
\begin{linenomath}
  \begin{equation} \label{eq_rho_Wood}
  \rho_w (\varphi)  = \rho_{\rm max} \exp ( - |\varphi / \sigma|^{\alpha} /2) \, .
  \end{equation}
  \end{linenomath}
The parameter $\sigma$ defines the azimuthal extension and $\alpha$ characterizes how flat the apex is \citep[see][]{Wood09b,Wood09c,Wood10,Wood11,Wood12}.  

  %{\S}{\bf --- Results of Wood for the bright front} \\
With a cylindrical rotation, Equation~(\ref{eq_rho_Wood}) defines a surface, and \citet{Wood09b} set a thin shell of density around it to model the bright front of ICMEs.   Then, \citeauthor{Wood09b} computed synthetic images by simulating the Thomson scattering.  The comparison with the observed COR and HI images of ICMEs observed by both STEREO spacecraft leads to the determination by visual inspection of the best values for $\sigma$ and $\alpha$.  
The results are ICME case-dependent with $\sigma$ in the interval $[25\ensuremath{^\circ} , 43\ensuremath{^\circ} ]$ and $\alpha$ in $[2, 3.3]$.

  %{\S}{\bf --- Results of Wood for the flux rope leading and trailing edges} \\
\citet{Wood09b} also used Equation~(\ref{eq_rho_Wood}) to define a density shell model around a flux rope, as follows.  They first defined the flux rope leading and trailing edges with Equation~(\ref{eq_rho_Wood}) and with two sets of $\sigma$ and $\alpha$ values.
Then, they defined a flux-rope like boundary with an elliptical (or circular) cross section passing through these two edges. 
Finally, they included a density layer around the cross-section boundary. The visual fit of the simulated brightness images with both STEREO data determines the best parameter values. The resulting $\sigma$ values are in the interval $[21\ensuremath{^\circ} , 40\ensuremath{^\circ} ]$ and the $\alpha$ values in $[2, 8]$.  \citeauthor{Wood09b} did not explicitly model the axis of the flux rope, but since its shape is between the leading and trailing edge shapes, we show below the same range of $\sigma$ and $\alpha$ parameters for the axis as quoted above.   
  
  %{\S}{\bf --- $\tan \lambda$, $\varphi_w (\lambda)$ for Wood} \\
In order to derive the statistical properties for the in situ data implied with the Wood's model, defined by Equation~(\ref{eq_rho_Wood}), we apply below the equations derived in Section~\ref{sect_spherical-geometry}. Then, $\tan \lambda$ is obtained from Equation~(\ref{eq_tanLambda}):
\begin{linenomath}
  \begin{equation} \label{eq_lA_Wood}
  \tan \lambda =\alpha \, \sigma^{-\alpha}~\varphi^{\alpha-1}/2 \, .
  \end{equation}
  \end{linenomath}
For the shape to not be sharp at the apex, \textit{i.e.} $\lambda =0$ for $\varphi =0$, the condition $\alpha >1$ should be satisfied. Next, since $\tan \lambda$ is a monotonously growing function of $\varphi$, the inversion of Equation~(\ref{eq_lA_Wood}) provides: 
\begin{linenomath}
  \begin{equation} \label{eq_phi_Wood}
  \varphi_w (\lambda) = 
  \left( \frac{2 \sigma^{\alpha} \tan \lambda}{\alpha} \right)^{1/(\alpha-1)} \, .
  \end{equation}
\end{linenomath}
  %{\S}{\bf --- $\rho_{\rm min}, $\varphi_{\rm max}$ and $\lambda_{\rm max}$ for Wood} \\
The profile of Equation~(\ref{eq_rho_Wood}) has no physical meaning for large $|\varphi|$ values,
\textit{i.e.} for very small $\rho$ values (as the shape would spiral around its origin).  We then set a minimum value $\rho > \rho_{\rm min}$, which implies $|\varphi| <\varphi_{\rm max}$ and $\lambda < \lambda_{\rm max}$ with:
  \begin{eqnarray} 
  \varphi_{\rm max}       &=&  \sigma ~\Big( 2\ln \frac{\rho_{\rm max}}{\rho_{\rm min}} \Big)^{1/\alpha}  
                             \, , \label{eq_phi_max_Wood} \\
  \lambda_{\rm max} &=& \tan^{-1} \Big( \alpha \, \sigma^{-\alpha}~\varphi_{\rm max}^{\alpha-1}/2  \Big)
                             \, . \label{eq_lA_max_Wood}
  \end{eqnarray}
In Figure~\ref{fig_wood} we selected $\rho_{\rm min}=0.2$ {AU} and $\rho_{\rm max}=1$ {AU}, which implied $\lambda_{\rm max}$ ranging from $74\ensuremath{^\circ}$ to $87\ensuremath{^\circ}$ for $\alpha$ ranging from 2 to 8. {The shape defined by this model, Equation~(\ref{eq_rho_Wood}), is shown in the left panel of Figure~\ref{fig_wood} with different colors for four values of $\alpha$ ($=2,4,6,8$). With the same colors, the corresponding probabilities of $\lambda$ for the axis and the shocks, Equation~(\ref{eq_Pl_Wood}) below, are shown in the central and right panels, respectively. These theoretical results are compared with the observed probability drawn with histograms for MCs analyzed by Lepping \& Wu and for shocks analyzed by Wang et al., respectively.}

\begin{figure}  %________________________ FIG ______________________________________    
\centering
\IfFileExists{2columnFigures.txt}{
\includegraphics[width=0.5\textwidth,clip]{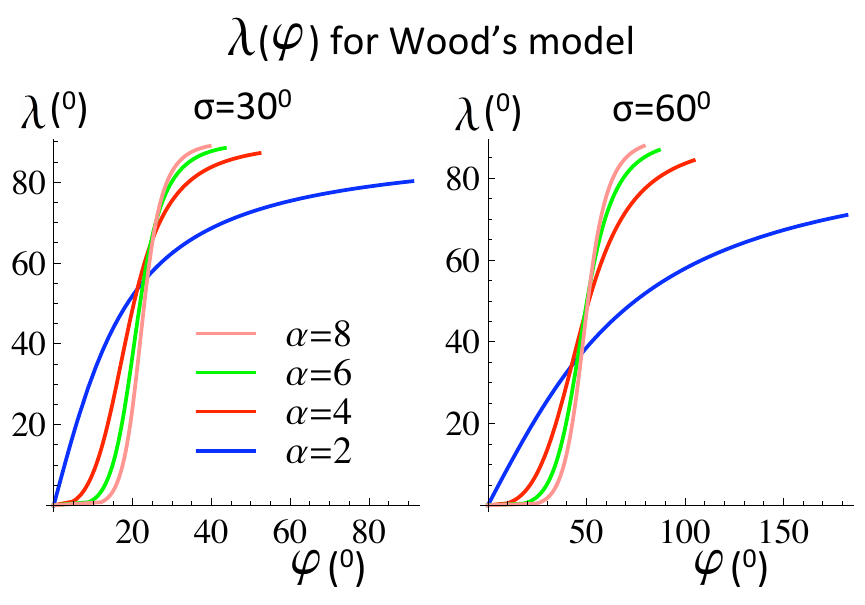}
  } {
\includegraphics[width=0.5\textwidth,clip]{fig_lambdaphi_wood}
}
\caption{Variation of the location angle $\lambda$ as a function of $\varphi$ for the Wood's model (Section~\ref{sect_Mod_Wood})
for four values of $\alpha$ and two values of $\sigma$. 
% All angles are in degree. 
The left panel has the same parameter values as in Figure~\ref{fig_wood}.
\m{These results are used in Section~\ref{sect_Mod_Wood} to analyze the incompatibility of Wood's model with the observed distributions of  $\lambda$ (Figure~\ref{fig_wood}).} 
}
\label{fig_lambda(phi)_wood}
\end{figure}

  %{\S}{\bf --- Properties of $\Pwl$} \\
The probability $\mathcal{P}(\lambda)$ is computed from Equation~(\ref{eq_Pl}), where $\varphi$ is replaced with Equation~(\ref{eq_phi_Wood}), so that the final expression is expressed as a function of $\lambda$ :
\begin{linenomath}
  \begin{equation} \label{eq_Pl_Wood}
  \mathcal{P}_{\rm w}(\lambda) = \frac{\mathcal{P}_{\varphi} (\varphi_w)}{\alpha-1} \, 
  \left( \frac{2 \sigma^{\alpha} }{\alpha} \sin^{2-\alpha} \lambda \; \cos^{-\alpha} \lambda               
         \right)^{1/(\alpha-1)} \, ,
  \end{equation}
  \end{linenomath}
with $\mathcal{P}_{\varphi} (\varphi_w)$ defined by Equation~(\ref{eq_pphi_axis}) {or Equation~(\ref{eq_pphi_shock}).}
Since $\alpha >1$, $\alpha/(\alpha-1) > 0$ and $\mathcal{P}_{\rm w}(\lambda)$ is always growing to infinity as $\lambda$ approaches $90\ensuremath{^\circ}$ (infinite branch).  If one restricts $\varphi$ to the interval $[0,\varphi_{\rm max}]$, $\lambda < \lambda_{\rm max} < 90\ensuremath{^\circ}$ so that the singularity disappears in $\mathcal{P}_{\rm w}(\lambda)$.  However, $\mathcal{P}_{\rm w}(\lambda)$ remains a sharply growing function of $\lambda$ for large $\lambda$ values, and this behavior starts at lower values of $\lambda$ for low values of $\alpha$ (see the {curves in the middle and right panels of} Figure~\ref{fig_wood}). $\mathcal{P}_{\rm w}(\lambda)$ is also singular at $\lambda=0$, again with an infinite branch, for $\alpha >2$ for the MC axis case (Figure~\ref{fig_wood}, middle panel) and for $\alpha > 3$ for the shock case (right panel).  This behavior is due to the relationship between $\lambda$ and $\varphi$ as shown in Figure~\ref{fig_lambda(phi)_wood} {for two $\sigma$ values (panels) and four $\alpha$ values (color curves)}.  For low and large values of $\lambda$, the curves are flat.  With a uniform distribution of $\varphi$, this implies a larger accumulation of cases with similar $\lambda$ where the curve is more horizontal, so a larger probability $\mathcal{P}(\lambda)$ for MC axis and shocks.

  %{\S}{\bf --- No match to any observations!} \\
The above properties of $\mathcal{P}_{\rm w}(\lambda)$ imply strong differences with the observed probabilities of the location angle for both the MC axis and the shock normal (Figure~\ref{fig_prob_lambda}). This is shown in Figure~\ref{fig_wood} {(middle and right panels)}, where the observed $\lambda$ probabilities for both MC axis and shocks, deduced from lists of MC/shock events, are put in the background for comparison (light blue histograms). The probability curves for the synthetic model of \citet{Wood09b} are computed with the parameter $\alpha$ in the range deduced from STEREO observations. A similar behavior is found for all $\sigma$ values, 
as illustrated with the evolution of $\lambda (\varphi )$ with two values of $\sigma$ in Figure~\ref{fig_lambda(phi)_wood}.  As $\sigma$ increases, $\varphi$ extends on a broader interval, while $\lambda (\varphi )$ keeps a similar shape implying similar $\mathcal{P}_{\rm w}(\lambda)$.

 As such, Wood's model, as described by Equation~(\ref{eq_rho_Wood}), is too {blunt} %flat 
around the apex, then too {tightly curved on the flanks} and again too flat in the legs/{flanks} to provide a satisfying evolution of $\mathcal{P}_{\rm w}(\lambda)$ comparable with in situ observations.  Even scanning the most appropriate range of $\alpha$ (within $[2,4]$), for which $\mathcal{P}_{\rm w}(\lambda)$ is less singular, does not provide a probability distribution comparable to any of the observed distributions (Figure~\ref{fig_prob_lambda}). 
Indeed, the differences between the model and the observations always remain of the order of the mean probability value (\rm{diff}\ $\approx 13$ to $33\times 10^{-3}$ with \rm{diff}\ defined by Equation~(\ref{eq_diff})).

  %{\S}{\bf --- No match to any observations!} \\
We conclude that the description of the axis and the shock shapes by Equation~(\ref{eq_rho_Wood}) does not provide any probability distribution compatible with the statistical in situ results.  At first sight, the shapes shown in the left panel of Figure~\ref{fig_wood} are not too far from the typically expected shape (\textit{e.g.} as shown in Figure~\ref{fig_schema}{b} 
and as observed by STEREO). However, $\mathcal{P}(\lambda)$ is a very sensitive function as it contains a second derivative of the shape (Equation~(\ref{eq_Pl})).  It implies that, even with limited statistics, the observed probabilities $\mathcal{P}(\lambda)$ provide strong constraints on the axis and shock shapes. 
 
\begin{figure*}  %________________________ FIG ______________________________________    
\centering
\includegraphics[width=\textwidth,clip]{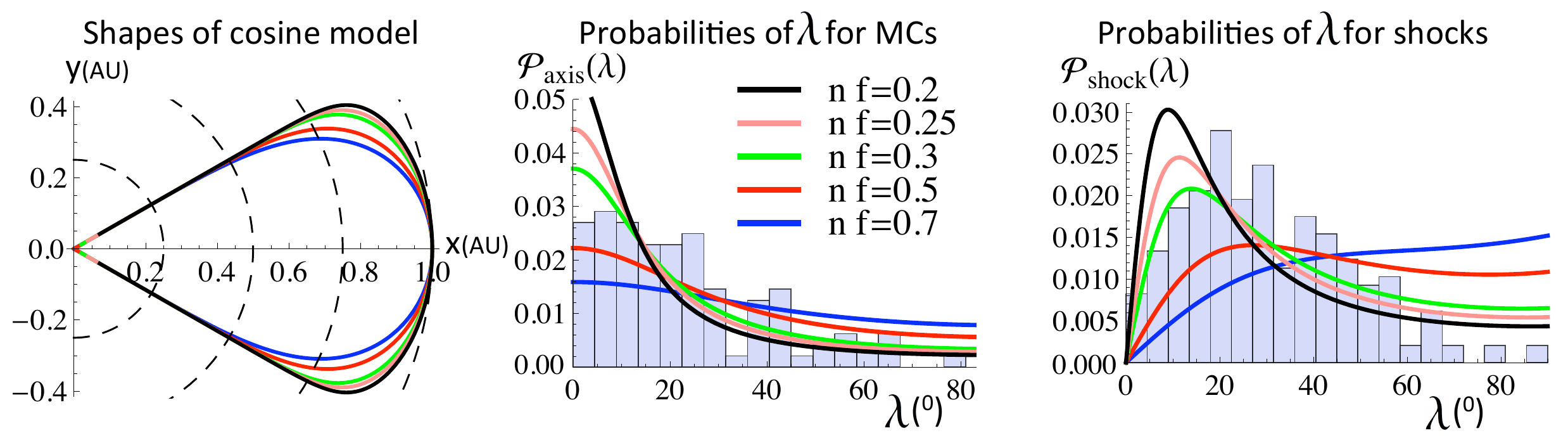}
\caption{Shapes and probabilities per unit $\lambda$, in degree, for the cosine model (Section~\ref{sect_Mod_cos}). The left panel shows the MC axis or the cross-section of the shock, as defined by Equation~(\ref{eq_rho_cos}), 
for different values of the $n\,f$ parameter.  The corresponding probabilities of $\lambda$ are shown in the {middle and} right panels together with, in the background, the observed probability drawn with histograms for MCs analyzed 
by Lepping \& Wu (middle panel) and for shocks analyzed by Wang et al. (right panel). \m{The colored lines show that the cosine model is the most consistent with the observed probability distributions for 
%value 
{the case with}
$n\,f \approx 0.3$.}
}
\label{fig_cos}
\end{figure*}

\subsection{Cosine model} %%%%%%%%
\label{sect_Mod_cos}

  %{\S}{\bf --- Definition of the model} \\
 Other analytical models can be used to study the properties of the statistical distributions of parameters for both shocks and MC axis. For example, a simple analytical model was introduced by \citet{Janvier14b} to describe the mean shape of shocks:
 \begin{linenomath}
  \begin{equation} \label{eq_rho_cos}
  \rho_c (\varphi)  = \rho_{\rm max} \cos ^n (f \varphi ) \, ,
  \end{equation}
  \end{linenomath}
with $f = 90\ensuremath{^\circ}/\varphi_{\rm max}$ so that $\rho(\varphi_{\rm max})=0$. This model can similarly describe the shape of the MC axis. In the following, we refer to this model as the cosine model, and all relevant parameters are denoted with a subscript {``c''}.  {The shape defined by the cosine model is shown in the left panel of Figure~\ref{fig_cos} for five values of the product $n\,f$}{, with $n\,f$ the product between $n$ and $f$, 
the most sensitive degree of freedom in this model.} 

For an apex fixed at a given distance, \textit{e.g.} ~1~AU (where the Wind and ACE spacecraft are located), this model has only two parameters (as the Wood's model): 
 \{$n$,$f$\} or equivalently \{$nf$,{$\varphi_{\rm max}$}\}.
Computing $\rm{d} \ln \rho_c / \rm{d}  \varphi$ and inverting Equation~(\ref{eq_tanLambda}) allows to express $ \varphi_c$ explicitly as a function of $\lambda$:
\begin{linenomath}
  \begin{equation} \label{eq_phi_cos}
  \varphi_c (\lambda) = \frac{1}{f} \tan^{-1} \left( \frac{\tan \lambda }{n\,f} \right) \, .
  \end{equation}
\end{linenomath}
  %{\S}{\bf --- Define $\Pcl$} \\
The probability $\mathcal{P}_{\rm c}(\lambda)$ of Equation~(\ref{eq_Pl}) is explicitly computed by including a second derivation of $\rm{d} \ln \rho_c / \rm{d}  \varphi$.  Eliminating $\varphi$ with Equation~(\ref{eq_phi_cos}), the expression for $\mathcal{P}_{\rm c}(\lambda)$ is rewritten as 
\begin{linenomath}
  \begin{equation} \label{eq_Pl_cos}
  \mathcal{P}_{\rm c}(\lambda) = \mathcal{P}_{\varphi} (\varphi_c) \, 
        \frac{n ~(1+\tan^2 \lambda)}{(n\,f)^2 +\tan^2 \lambda} \, .
  \end{equation}
  \end{linenomath}
With $\varphi_c$ expressed with Equation~(\ref{eq_phi_cos}), $\mathcal{P}_{\rm c}(\lambda)$ becomes an explicit function of $\lambda$. 

  %{\S}{\bf --- Behavior of $\mathcal{P}_{\rm c}(\lambda)$} \\
The probability $\mathcal{P}_{\rm c}(\lambda)$ is strongly dependent on the product $n\,f$ as shown in Figure~\ref{fig_cos} {for the axis and the shocks in the central and right panels, respectively, with the same color convention than for the shapes (left panel)}.  Values of $n\,f \geq 0.5$ \m{give} $\mathcal{P}_{\rm c}(\lambda)$ functions that are too flat compared with the observed probabilities (as reported in the background for both MC axis and shock normal). This is especially true for the shocks, since the observed decrease of $\mathcal{P}(\lambda)$ for large $\lambda$ is not reproduced {for $n\,f \geq 0.5$} (e.g., the blue curve for $n\,f = 0.7$ is monotonously increasing).  On the contrary, values $n\,f \leq 0.25$ imply that the $\mathcal{P}_{\rm c}(\lambda)$ functions  are too peaked near the origin compared with the observations (e.g., the black curve for $n\,f = 0.2$).  {Furthermore, we found the same constraints on $n\,f$ by comparing the cosine model to} 
 %More generally, very similar results are obtained by comparing with all 
the other observed probabilities (\textit{e.g.} those shown in Figure~\ref{fig_prob_lambda}).
It implies that the $n\,f$ values compatible with observations are located in a narrow interval $[0.25,0.5]$. This corresponds to a {well-constrained} shape as shown in the left panel of Figure~\ref{fig_cos}, in between red and pink shapes, because a large modification of $\mathcal{P}_{\rm c}(\lambda)$ implies only a small deformation of the shape.

  %{\S}{\bf --- Best fit of $\mathcal{P}_{\rm c}(\lambda)$} \\
We quantify the above results by minimizing the least square difference between the observation and model distributions, so minimizing the function ``diff'' defined in Equation~(\ref{eq_diff}) for each observed probability. This provides the best parameters $n\,f$ and $\varphi_{\rm max}$. In fact,
$\varphi_{\rm max}$ has a very weak effect on $\mathcal{P}_{\rm c}(\lambda)$ \citep{Janvier14b} so that only the best $n\,f$ value is determined.   The results are summarized in Table~\ref{Tab_diff}.  
For MC axis, $n\,f$ is almost the same, $\approx 0.35$, for the data of \citet{Lepping10} and \citet{Feng10}, while $n\,f$ is slightly larger, $\approx 0.5$, for \citet{Lynch05} because its $\mathcal{P}(\lambda)$ has a larger tail for large $\lambda$ values (Figure~\ref{fig_prob_lambda}). For shocks, all the results are clustered around $n\,f = 0.3$.   

  %{\S}{\bf --- Conclusion} \\
The above results imply that the deduced mean shape of MC axis is nearly independent of the data set selected.  This includes partly different MCs (see Section~\ref{sect_Precision}), and more importantly different input in the analyzed procedure from different authors \citep[\textit{e.g.} the MC axis depends critically on the author's choice of the MC boundaries,][]{Dasso06}.   The deduced mean shape of shocks is also very close from the one deduced from the data sets of \citet{Feng10} and \citet{Wang10} while {the determined {shock} normals have significant differences for the same analyzed shock (Section~\ref{sect_Precision}). }
  
\begin{table}  %________________________ TABLE ______________________________________
   \caption{ Best fitted cosine and ellipsoidal models to various data sets.}
   \label{Tab_diff}
    \begin{tabular}{@{}lr|c@{~~}c|c@{~~}c@{}}    
\hline
\hline
\multicolumn{2}{c|}{Observations} &\multicolumn{2}{c|}{cos. mod.}&\multicolumn{2}{c}{ellip. mod.}\\
data set & $N_{\rm case}$ &$n\,f$ &$\rm{diff}^{\mathrm{~a}}$&$b/a$&$\rm{diff}^{\mathrm{~a}}$ \\
\hline
\multicolumn{5}{c}{results with \bf MC axis}\\
\hline
Lepping \& Wu, all         &107 & 0.36 & 4.5 & 1.28 & 4.1 \\
Lepping \& Wu, {quality 1,2$^{\mathrm{~b}}$}& 74 & 0.36 & 6.9 & 1.25 & 6.4 \\
Lynch \textit{et al.}                &132 & 0.50 & 5.0 & 1.10 & 4.4 \\
Feng \textit{et al.} , axis          & 62 & 0.34 & 4.8 & 1.29 & 4.5 \\
\hline
\multicolumn{5}{c}{results with \bf shock normal}\\
\hline
Feng \textit{et al.} , shock         & 62 & 0.40 & 7.8 & 1.21 & 6.9 \\
Wang , all                &216 & 0.29 & 4.5 & 1.39 & 3.7 \\
Wang , \m{ICME not detected}$^{\mathrm{~c}}$ & 99 & 0.32 & 6.0 & 1.32 & 5.1 \\
Wang , all ICME           &117 & 0.26 & 4.4 & 1.45 & 3.6 \\
Wang , non-flux rope ICME & 36 & 0.23 & 6.1 & 1.53 & 5.8 \\
Wang , MC-like            & 36 & 0.23 & 6.5 & 1.57 & 5.8 \\
Wang , MC                 & 45 & 0.32 & 7.8 & 1.36 & 7.3 \\
\hline
    \end{tabular}
\begin{list}{}{}
\item[$^{\mathrm{a}}$] The difference computed with Equation~(\ref{eq_diff}) and multiplied by $1000$. 
\item[$^{\mathrm{b}}$] {The quality is defined in \citet{Lepping90} according to the $\chi^2$ value of the fit of a flux-rope model to data. Here the two best groups are used.}
\item[$^{\mathrm{c}}$] {Shocks not followed by an ICME. Most of them are thought to be shock flanks which extend beyond the ICME \citep[see][]{Janvier14b}.}
\end{list}
\end{table}

\begin{figure}[t!]    
\centering
\IfFileExists{2columnFigures.txt}{
\includegraphics[width=0.5\textwidth,clip]{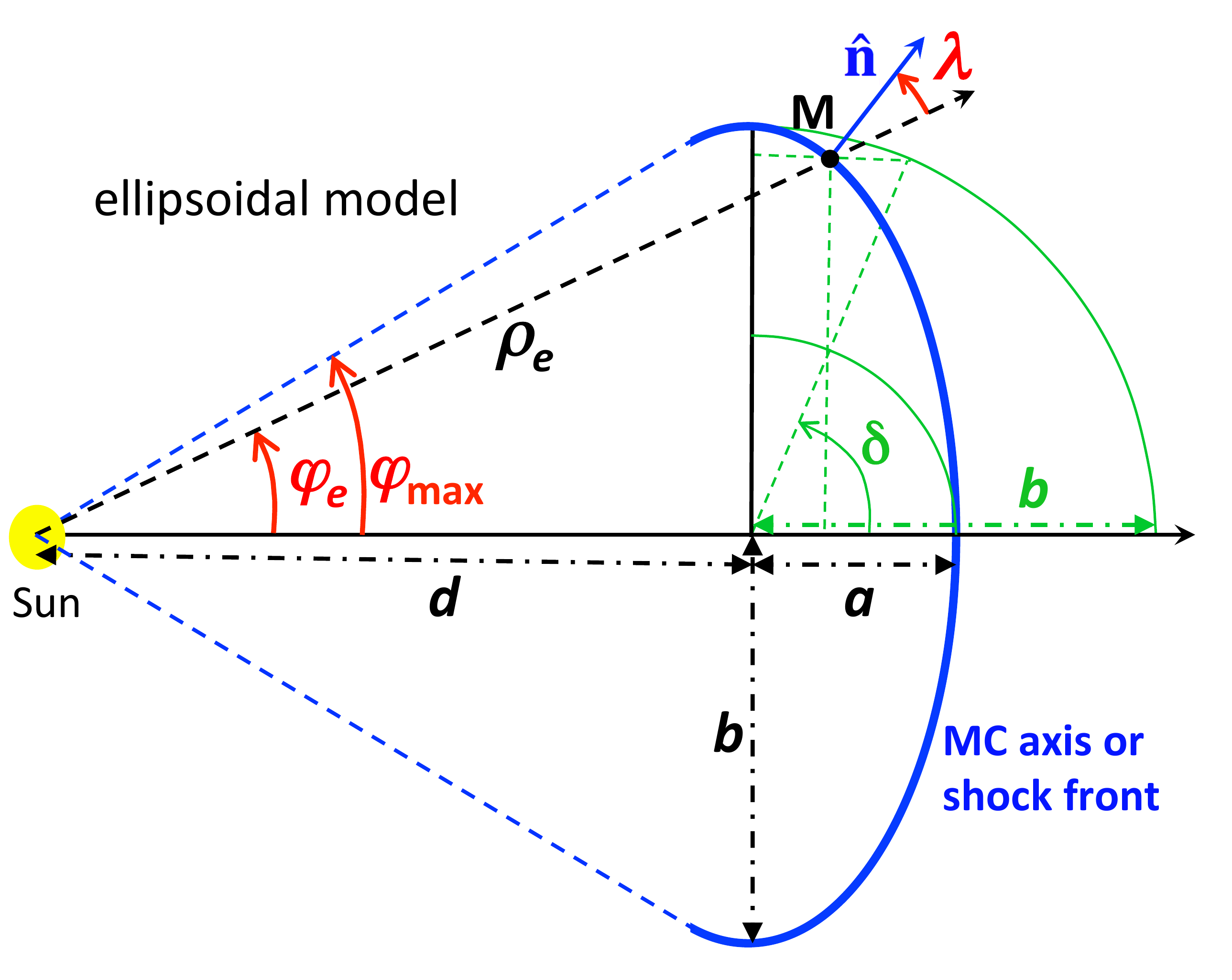}
  } {
\includegraphics[width=0.6\textwidth,clip]{fig_schema_ellipsoidal}
}
\caption{{Diagram defining the elliptical model for the flux-rope axis or shock shape.
M is the point of interest (where the spacecraft crosses the structure).
$\varphi_e$ is the angle of the cylindrical coordinates ($\rho_e,\varphi_e$). The location angle $\lambda$ is defined between the normal and the local radial from the Sun.
The maximum angular extension $\varphi_{\rm max}$ is outlined by radial segments tangent to the ellipse. $\delta$ is the angle used to parametrized the ellipse.
} }
 \label{fig_schema_ellipsoidal}
\end{figure}

\subsection{Ellipsoidal model} %%%%%%%%
\label{sect_Mod_ellip}

  %{\S}{\bf --- Definition of the model} \\
 The ellipsoidal model was introduced by \citet{Janvier13} to describe the mean shape of the MC axis. Its derivation is less simple than for the above cosine model, although $\mathcal{P}(\lambda)$ can still be derived analytically. The MC axis is described by an ellipse of half size $a$ and $b$ in the radial and orthoradial directions, respectively {(Figure~\ref{fig_schema_ellipsoidal})}.  The ellipse centre is at a distance $d$ from the Sun, and we define a point M on the axis where the spacecraft crosses the structure, situated at a distance $\rho_e$ from the Sun:  
 \begin{linenomath}
  \begin{equation} \label{eq_rho_ellip}
  \rho_e = \sqrt{(d+a \cos \delta)^2 +  (b \sin \delta)^2} \, ,
  \end{equation}
  \end{linenomath}
where $\delta$ is the angle defining the position of M from the ellipse centre.  The azimuthal coordinate, $\varphi_e$, writes
\begin{linenomath}
   \begin{equation} \label{eq_phi_ellip}
  \tan \varphi_e = b  \sin \delta /(d+a \cos \delta) \, .
  \end{equation}
  \end{linenomath}
These equations, although first defining the MC axis, can similarly describe the shock shape (considering a surface symmetric around the Sun-apex line) as a parametric curve $\big(\rho_e (\delta), \varphi_e (\delta)\big)$.
The maximum angular extension of the model, $\varphi_{\rm max}$, is defined by:
\begin{linenomath}
  \begin{equation} \label{eq_phimax_ellip}
  \tan \varphi_{\rm max} = b / \sqrt{d^2-a^2}  \, .
  \end{equation}
  \end{linenomath}
For an apex fixed at a given distance, this model has only two parameters similarly to the two previous models: $a/d$ and $b/d$, or equivalently $b/a$ and $\varphi_{\rm max}$.

  %{\S}{\bf --- Define $\varphi_c (\lambda)$} \\
The location angle $\lambda$ is expressed with Equation~(\ref{eq_tanLambda}) as:
 \begin{linenomath}
  \begin{equation} \label{eq_tanLambda_ellip}
  \tan \lambda = \frac{a \sin \delta \cos \varphi_e - b \cos \delta \sin \varphi_e}
                  {a \sin \delta \sin \varphi_e + b \cos \delta \cos \varphi_e} \, .
  \end{equation}
  \end{linenomath}
Both  $\varphi_e(\delta )$ and $\lambda (\delta )$ are monotonously growing functions of $\delta$, then $\varphi_e$ is an implicit function of $\lambda$.  

  %{\S}{\bf --- Define $\mathcal{P}_{\rm c}(\lambda)$} \\
The derivation of the probability $\mathcal{P}_{\rm e}(\lambda)$ requires several computation steps as outlined in 
\citet{Janvier13}. The result is:
 \begin{linenomath}
  \begin{equation} \label{eq_Pl_ellip}
  \mathcal{P}_{\rm e}(\lambda) = \mathcal{P}_{\varphi} (\varphi_e) \, / \, |\rm{d} \lambda/\rm{d} \varphi | \, ,
  \end{equation}
    \end{linenomath}
with 
 \begin{linenomath}
  \begin{eqnarray} 
  \IfFileExists{2columnFigures.txt}{
  \frac{\rm{d} \lambda}{\rm{d} \varphi} = -1 
  &+& \frac{1+\tan^2 \delta}{1+ (a/b)^2\tan^2 \delta} \,
    \frac{a}{b \cos \varphi_e}  \nonumber \\
  && \times \frac{d+a \cos \delta}{a \sin \delta \sin \varphi_e + b \cos \delta \cos \varphi_e}
      \, .\nonumber
  } {
  \frac{\rm{d} \lambda}{\rm{d} \varphi} = -1 
  &+& \frac{1+\tan^2 \delta}{1+ (a/b)^2\tan^2 \delta} \,\,
    \frac{a}{b \cos \varphi_e} \,\,  
   \frac{d+a \cos \delta}{a \sin \delta \sin \varphi_e + b \cos \delta \cos \varphi_e}
      \, .\nonumber
}
  \end{eqnarray}
    \end{linenomath}
With Equations \ref{eq_phi_ellip} and \ref{eq_tanLambda_ellip} {providing the monotonous functions $\varphi_e(\lambda )$ and $\delta  (\lambda)$,} $\mathcal{P}_{\rm e}(\lambda)$ is an implicit single-value function of $\lambda$.

\begin{figure*}  %________________________ FIG ______________________________________    
\centering
\includegraphics[width=\textwidth,clip]{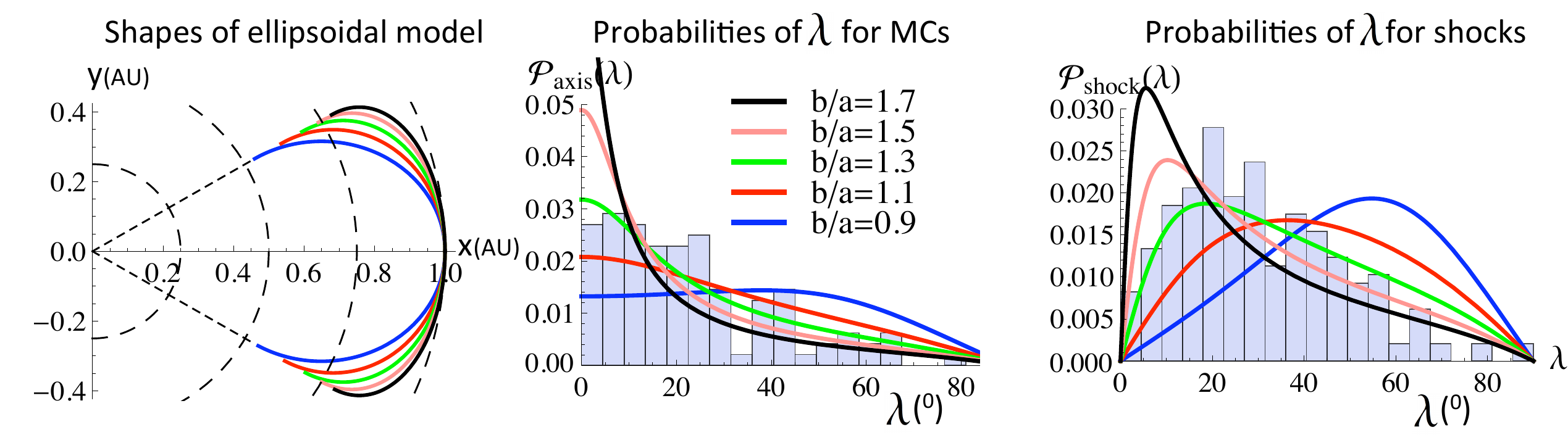}
\caption{Shapes and probabilities per unit $\lambda$, {in degree,} for the ellipsoidal model (Section~\ref{sect_Mod_ellip}). {We added a straight sunward edge to the ellipsoidal shape (dashed line) to visualize the maximum extension $\pm \varphi_{\rm max}$ (defined in Figure~\ref{fig_schema}b).} The left panel shows the MC axis or the cross-section of the shock, as defined by Equation~(\ref{eq_rho_ellip}), for different values of the aspect ratio $b/a$ parameter.  The corresponding probabilities of $\lambda$ are shown in the middle and right panels together with, in the background, the observed probability drawn with histograms for MCs analyzed by Lepping \& Wu (middle panel) and for shocks analyzed by Wang et al. (right panel). \m{The comparison between the colored lines and the histograms shows that the ellipsoidal model with $b/a \approx 1.3$ best fits with the observed data.}
}
\label{fig_ellipsoid}
\end{figure*}

   %{\S}{\bf --- Behavior of $\mathcal{P}_{\rm e}(\lambda)$} \\
We {show in the left panel of} Figure~\ref{fig_ellipsoid} the different shapes obtained with such a model for different aspect ratios. We also {show} the associated probability functions (colored curves) on top of histograms of the observed distributions of $\lambda$ for both MC axis (middle panel) and shock normal (right panel), {with the same drawing convention than in Figure~\ref{fig_cos}}. 
For both MC axis and shock normals, the probability function $\mathcal{P}_{\rm e}(\lambda)$ matches the observed $\mathcal{P}(\lambda)$ the best for a value $b/a \approx 1.3$ (green curve). This corresponds to an ellipsoidal shape slightly elongated in the orthoradial direction.  Although the parametrised curve does not change much for nearby values, the probability $\mathcal{P}_{\rm e}(\lambda)$ is very sensitive to the parameter $b/a$ as there are large differences from the observed $\mathcal{P}(\lambda)$ for these nearby values, \textit{e.g.} $b/a=1.1$ and $1.5$ (red and pink curves respectively in Figure~\ref{fig_ellipsoid}). 
Compared to the cosine model, the ellipsoidal model provides an even closer match with the observations, especially for large $\lambda$ values of the $\mathcal{P}(\lambda)$ function for shock normals (\textit{e.g.}, one can compare the green curves in Figures \ref{fig_cos} and \ref{fig_ellipsoid}). 

\begin{figure*}  %________________________ FIG ______________________________________  
\centering
\includegraphics[width=\textwidth,clip]{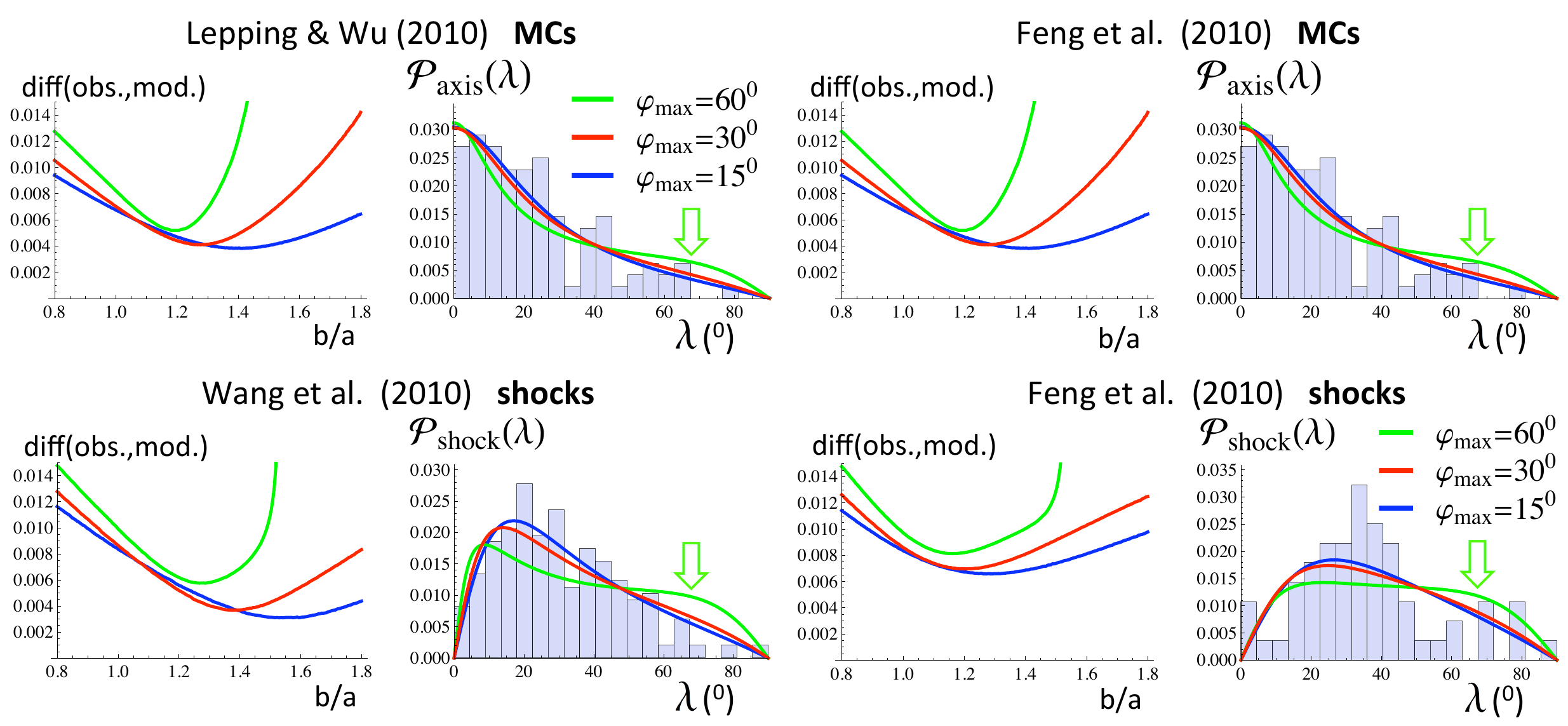}
\caption{Minimisation of the difference between the observations and the ellipsoidal model (Section~\ref{sect_Mod_ellip}).
The top row shows results for the MC axis using two different samples of MCs, while the bottom row shows shock normal results using two samples of shocks.
The difference function is defined by Equation~(\ref{eq_diff}) and it is plotted as a function of the aspect ratio $b/a$ in the left-hand plot of each pair. The results are shown for three values of $\varphi_{\rm max}$ (equal to $\varphi_{\rm max, axis}$ for MCs and to $\varphi_{\rm max, shock}$ for shocks as defined in Figure~\ref{fig_schema}{b}). In the right-hand plot of each pair, the probabilities of the best models (\textit{i.e.} minimising \rm{diff}\ of Equation~(\ref{eq_diff})) are shown together with, in the background, the observed probability drawn with histograms. The green arrows show how the tail increases with $\varphi_{\rm max} = 60 \ensuremath{^\circ}$.  \m{The main result is that the deduced shape, characterized by  $b/a$ minimizing the difference function, is almost independent of $\varphi_{\rm max}$ for low values ($< 60 \ensuremath{^\circ}$) while for larger $\varphi_{\rm max}$ value the ellipsoidal model is further away from observations.} %All angles are in degree.} 
}
\label{fig_diff}
\end{figure*}

  %{\S}{\bf --- Best fit of $\mathcal{P}_{\rm e}(\lambda)$} \\
We find the best ellipsoidal description of the MC axis and shock {front} by using the least square difference technique (as in Section~\ref{sect_Mod_cos}).
In Figure~\ref{fig_diff}, we show how to obtain the best fits for two different samples of MCs and shocks {with a pair of plots for each sample. We also show three  values of $\varphi_{\rm max}$ (color curves)}.
For each sample, we show in the left panel {of each pair} the evolution of the $\rm{diff}$ value, {Equation~(\ref{eq_diff}),} as a function of the ellipse aspect ratio $b/a$. The lowest values of $\rm{diff}$ correspond to the best fits, and the associated probability functions are then reported on top of the observed probability distribution in the right panel {of each pair}.

The minimum $\rm{diff}$ values are all similar for low $\varphi_{\rm max}$ values (\textit{e.g.} $15 \ensuremath{^\circ}$ and $30 \ensuremath{^\circ}$) for both MCs and shocks (Figure~\ref{fig_diff}), while they increase for larger $\varphi_{\rm max}$ values (\textit{e.g.} $60 \ensuremath{^\circ}$).
{Indeed the probability of $\lambda$ is weakly dependent of $\varphi_{\rm max}$ for low $\varphi_{\rm max}$ values \citep{Janvier13,Janvier14b}, then ICMEs with variable $\varphi_{\rm max}$ can be analyzed together.
However,} the probability function evolution, for large $\varphi_{\rm max}$ values, $\mathcal{P}_{\rm e}(\lambda)$ has a too large tail for $\lambda \ge 60 \ensuremath{^\circ}$ values compared with observations (as shown with the green arrows in Figure~\ref{fig_diff}), especially for shocks (bottom panels, {right panel of each pair}). As such, {within the limits of small statistics for large $\lambda$ values}, large values of $\varphi_{\rm max}$ are not consistent with the observed probability functions, 
which indicates that ICME shocks are not typically so spatially broad.  {This is in agreement with 
the predominance of solar sources of near-Earth 
ICMEs close to central meridian \citep[64\% are within $20\ensuremath{^\circ}$ of central meridian][]{Richardson10}
as well as with} a mean half angular extension of CMEs of $30 \ensuremath{^\circ}$ estimated from coronagraphic observations from CMEs launched close to the solar limb to minimize the projection effects \citep[][]{Wang11}.

  %{\S}{\bf --- Conclusion for the ellipsoidal model} \\
The best $b/a$ values are summarized in Table~\ref{Tab_diff} for $\varphi_{\rm max}=30 \ensuremath{^\circ}$.  In all cases a lower minimum $\rm{diff}$ value is found for the ellipsoidal model, confirming the conclusion drawn above from Figures \ref{fig_cos} and \ref{fig_ellipsoid} that the ellipsoidal model is a better fit than the cosine model. For MC axis, $b/a$ is close to $1.3$ except for the data of \citet{Lynch05} with $b/a \approx 1.1$, 
indicating a slightly more bent shape.
For the shock normal, the value of $b/a$ that fits the best with observations is around $1.4$, except for the data of \citet{Feng10} for which $b/a \approx 1.2$, indicating also a slightly more bent shape.  
{These variations of $b/a$ are comparable to the uncertainty of $b/a \approx \pm 0.1$} for both MC axis and shock normals as indicated by the minimum region extension of {the $\rm{diff}$ function} (Figure~\ref{fig_diff}).

  %{\S}{\bf --- General conclusion} \\
More globally, a comparable shape is expected for the MC axis and the shock surface as the flux rope is only separated from the shock by the sheath (Figure~\ref{fig_schema}{b}). This is indeed what is found here with both cosine and ellipsoidal models, when their $\mathcal{P}(\lambda)$ is fitted to observations (Figures \ref{fig_cos} and \ref{fig_ellipsoid}). This is quite a remarkable result, considering that 1) the in situ data sets and 2) the techniques to find the MC axis and shock normal are all different. These comparable results from different types of data provide a strong case for the shape determined by cross-validating the results.

\subsection{Robustness of the derived axis and shock shapes} %%%%%%%%
\label{sect_Global_Robustness}

% {\S}{\bf --- Problem: large errors in data!} \\
The $\lambda$ values deduced for the different samples of MCs and shocks, and analyzed by different authors, show a large dispersion 
(Section~\ref{sect_Precision}, Figure~\ref{fig_correlation}). This seems a priori incompatible with the results above, where similar axis and shock shapes are derived from these various data sets (Sections \ref{sect_Mod_cos} and \ref{sect_Mod_ellip}, 
Table~\ref{Tab_diff}). To understand these results, \m{we investigate the effect of the $\lambda$ error level in the Appendix (Section~\ref{sect_Appendix})} for both the axis and the shock normal on the determined shapes. 
 
% {\S}{\bf --- Conclusion: shape stable  } \\
{We investigate the implications of errors on $\mathcal{P}_{\rm obs}(\lambda)$, then on the deduced MC axis and shock shapes. 
We conclude that, while there are large fluctuations in the $\lambda$ estimations by different authors for the same events (Figure~\ref{fig_correlation}), the observed probability distributions $\mathcal{P}_{\rm obs}(\lambda)$ are less affected by these errors because of the averaging implied when building an histogram.  Furthermore, the deduced axis and shock shapes are even much less affected by these errors because they depend only on the global properties of the histograms.  Indeed, increasing the estimated dispersion, \m{$\sigma_{\rm c}$,} by up to a factor three only has a weak effect on the deduced shapes (Figure~\ref{fig_convol}, {left panels, the same color being used for the same case on corresponding left and right panels of a panel pair}).
Moreover, the deconvolution of $\mathcal{P}_{\rm obs}(\lambda)$ by a Gaussian kernel sets an upper limit to the standard deviation of the \m{$\lambda$ errors}, up to $\approx 18 \ensuremath{^\circ}$ for MC axis and $\approx 8 \ensuremath{^\circ}$ for shock normals.   Within such limits, the deduced MC axis and shock shapes are not significantly influenced by the error level \m{on $\lambda$.}}

% {\S}{\bf --- Differrence with Feng et al.  } \\
\m{The above results contrast} with the results of \citet{Feng10} who compared the MC axis and shock normal directions directly with each MC-shock pair.   The large errors shown in Figure~\ref{fig_correlation} do not allow to derive meaningful results from this direct approach. 
However, with the same data set, we can perform a statistical analysis and derive the generic MC axis and shock shape as shown above.

%%%%%%%%%%%%%%%%%%%%%%%%%%%%%%%%%%%%%%%%%%%%%%%%%%%%%%%%%%%%%%%%%%%%%%%%%%%%%%%%%%%%%
\section{{Summary of the main results}} %%%%%%%%%%%%%%%%%%%%%%%%%%%%%
\label{sect_Summary}

%   {\S\bf General summary and main aim of the paper} \\
In the present paper, we propose an original statistical study based on different catalogues of flux ropes and shocks observed in situ. {We} compare different analytical models to derive and quantify the most probable generic flux rope axis and shock shape.
While in situ data provide only local information along the spacecraft trajectory crossing a MC/shock, our method combines the information from large sets of events to statistically derive global information on their generic shape.

%   {\S\bf Compare different author results} \\
Our study is based on a series of papers reporting fitted parameters associated with flux ropes inside magnetic clouds \citep{Lynch05,Feng10,Lepping10} as well as parameters associated with properties of shocks driven by ICMEs \citep{Feng10, Wang10}. While the FR fits are all made with the Lundquist model, and the shocks with the MHD Rankine-Hugoniot relations, differences in their application 
% and more importantly in the definition of the MC boundaries or the data regions used around shocks, 
lead to a dispersion \m{of the deduced parameters} for the same events (Section~\ref{sect_Precision}, Figure~\ref{fig_correlation}). 
This dispersion is \m{more important} for the location angle $\lambda$, which defines the location of the spacecraft crossing the interplanetary structure (see Figure~\ref{fig_schema}{b} for the definition of $\lambda$).
%Despite the significant dispersion found in a 'case by case' comparison (Figure~\ref{fig_correlation}), due to the fact that there is not a systematic underestimation/overestimation of one method/author when compared with another one, 
\m{However, the} probability distributions of observed $\lambda$ have comparable behavior \m{for all data sets} (decreasing function for the MC axis, and a Gaussian-like distribution peaking around $\lambda \sim 30\ensuremath{^\circ}$ for shock normals, see Figure~\ref{fig_prob_lambda}).

%   {\S\bf Fit to models} \\
We then compared the observed distributions with those obtained from three synthetic models for the MC axis and shock {fronts}: the Wood's (from \citealt{Wood09}), cosine \citep{Janvier14b} and ellipsoidal \citep{Janvier13} models. 
%P All these models can be expressed analytically in term of the two spherical coordinates $\rho$ and $\varphi$, themselves expressed with the location angle $\lambda$.  Since in a previous work \citep{Janvier14c} we have found only a weak asymmetry between the MC legs by deriving the axis shape from direct integration of the data, we consider here only symmetric axis models.  For the shocks, we suppose a cylindrical symmetry around the line Sun-apex.  Then, 
We investigated the different analytical distribution functions $\mathcal{P}(\lambda)$ for each model, by varying the shape of the MC axis/shock. 
%This comparison is first made by plotting the model distributions $\mathcal{P}(\lambda)$, obtained by varying the shape parameters, on top of the observed distributions $\mathcal{P}_{\rm obs}(\lambda)$ (see Figures \ref{fig_wood}, \ref{fig_cos} and \ref{fig_ellipsoid}).
We scanned the range of possible parameter {values} for each model and we determine the best model parameters by computing the difference between the distribution function $\mathcal{P}(\lambda)$ with each observed distribution $\mathcal{P}_{\rm obs}(\lambda)$.
%, see Equation~(\ref{eq_diff}). 

%   {\S\bf Results for Wood's model} \\
 The Wood's model was developed to analyze imager data \m{of CMEs} obtained from three view points (the two STEREO and SOHO spacecraft). % for some CMEs and to derive their travel properties (speed and direction).
While this model provides axis/shock shapes which at first look plausible, its derived probability function of $\lambda$, $\mathcal{P}_{\rm w}(\lambda)$, is incompatible with all the observed distributions, $\mathcal{P}_{\rm obs}(\lambda)$, of MC axis and shock for all the range of parameters derived by fitting this model to imager data  %. The differences between $\mathcal{P}_{\rm w}(\lambda)$ and $\mathcal{P}_{\rm obs}(\lambda)$ are important for all parameter values 
(Figure~\ref{fig_wood}). Also, this model has a too flat shape at the apex, which would indicate that most magnetic clouds and shocks would have been crossed at a low $\lambda$, which is not the case.  Then, these differences are intrinsic to the model rather than to the specific CME cases studied with the imaging instruments, and cannot also be accounted as the result of an evolution of the shape from the inner heliosphere to 1 AU.
      
 %  {\S\bf Best shapes} \\
By contrast to the Wood's model, both the cosine and ellipsoidal models are able to reproduce the observations, $\mathcal{P}_{\rm obs}(\lambda)$, within a narrow range of parameter values (Figures \ref{fig_cos} and \ref{fig_ellipsoid}).  Still, the ellipsoidal model provides the best fit for both the $\lambda$ distribution for MC axis and shock normals for all data sets (Section~\ref{sect_Mod_ellip}). 
The best ellipsoidal shape is obtained for an aspect ratio $b/a \sim 1.2$ for the MC axis and $b/a \sim 1.3$ for the shock normal. % (Figure~\ref{fig_diff} and Table~\ref{Tab_diff}).
This is quite a remarkable result, first because it allows us to define the quantitative generic shape of the MC axis and the shock front, and second because it shows that both structures have a similar shape, \m{while MC and shock data are independent.}
% with a shock less bent than the MC axis, as expected if the faster propagation of a MC than the ambient solar wind is the origin of the shock.
%Although the maximum angular extension $\varphi_{\rm max}$ remains a free parameter with the studied models, 
%We also show that the values of the maximum angular extension $\varphi_{\rm max}$ cannot be too large ($ < 60\ensuremath{^\circ}$, Figure~\ref{fig_diff} {within the limits of the statistical fluctuations}). Finally, 

\m{Moreover,} although the observed probability distributions $\mathcal{P}_{\rm obs}(\lambda)$ have some differences from one sample of MC/shock to another (Figure~\ref{fig_prob_lambda}), the results of the ellipsoidal and cosine models are close for different data sets (Figure~\ref{fig_diff} and Table~\ref{Tab_diff}), confirming that the derived shapes of axis/shock is weakly dependent on the details of $\mathcal{P}_{\rm obs}(\lambda)$. 
 %, in particular on the statistical fluctuations and the biases present on $\lambda$. 
Indeed, for any model, $\mathcal{P}(\lambda)$ is a very sensitive function of the axis/shock shape as it contains a second derivative of the shape (Equation~(\ref{eq_Pl})).  This implies that the global behavior of $\mathcal{P}(\lambda)$ defines precisely the generic shape of MC axis/shock.  

%   {\S\bf Effect of errors on $\lambda$} \\
We investigated the reason for such weak dependence on \m{the MC axis/shock shapes, by analyzing} the effect of statistical errors introduced in $\mathcal{P}_{\rm obs}(\lambda)$. % (Section~\ref{sect_Appendix}). 
%For each event, a bias is introduced on $\lambda$, but for a large set of events, by comparing the results of different authors, the errors appear closer to a random distribution (Figure~\ref{fig_correlation}).  We approximate this distribution by a Gaussian distribution with the same standard deviation (see Equation~\ref{eq_kernel}, Figure~\ref{fig_diff_lambda}). Then,  we 'cleaned' the observed distributions of $\lambda$ for both MC axis and shocks, and included different error distributions to investigate the changes introduced by the errors of $\lambda$ on the deduced shapes.  
We found in particular that different levels of errors introduced moderate modifications of the probability distributions while the modifications of the axis/shock shape are minor (Figure~\ref{fig_convol}).
Indeed, a significant change of the shapes would require a large modification of $\mathcal{P}_{\rm obs}(\lambda)$,
as shown in Figs.~\ref{fig_wood}, \ref{fig_cos} and \ref{fig_ellipsoid}.  We conclude that the observed distributions $\mathcal{P}_{\rm obs}(\lambda)$ set a strong constraint on the generic axis/shock shapes. 
   
%%%%%%%%%%%%%%%%%%%%%%%%%%%%%%%%%%%%%%%%%%%%%%%%%%%%%%%%%%%%%%%%%%%%%%%%%%%%%%%%%%%%%
\section{{Conclusions and Implications}} %%%%%%%%%%%%%%%%%%%%%%%%%%%%%
\label{sect_Conclusion}

 %  {\S\bf Summary} \\
 \m{We have derived the generic shape of MC axis and ICME shocks from different published catalogues of events computing the MC axis or shock normal from in situ data.  
The MC fits are all made with the Lundquist model, and the shocks are analyzed with the MHD Rankine-Hugoniot relations. Both methods have their own limitations.  
However, these catalogues have presently the largest number of studied cases compared with other techniques.  
Moreover, our statistical methods developed to deduced the shapes of these structures can {be applied} to any other data samples that provide the MC axis and/or the shock normals.  This is a further motivation to extend other catalogues.}   
  
Deriving generic shapes for the MC axis and shocks has several implications. 
 %  {\S\bf Implications for imager studies} \\
First, they could be used in analysis of imager data with single viewpoint or stereoscopic observations, as follows.  Presently the imager data of ICME require a model in order to derive physical properties such as the velocity and the direction of propagation. Past studies were derived with simple models (the CME front is either supposed to have a negligible extension, or to be spherical around the Sun or attached to it, or to be described by Wood's model). 
However, the results are sensitive to the model selected \citep[\textit{e.g.}][]{Mostl14}. We propose the ellipsoidal model, derived from \m{our statistical results},
%the best fit to in situ data, 
as an alternative model to analyze imager data. 
%\m{This conclusion comes from our comparative study of different samples of MCs and shocks with novel statistical techniques.  
\m{Nonetheless, with this elliptical model, it is possible to generalize} the equations derived by \citet{Davies13} for a detached spherical model to an ellipsoidal model of the ICME front.   Introducing a more elaborated model implies more free parameters which are not necessarily constrained by imager data. However, the in situ data constrain well the aspect ratio $b/a$ of the model to a narrow range of values, then they provide a reference case for application to imager data.

%   {\S\bf Implications for forecast} \\
A second implication of our results is for forecasting ICME arrival times at Earth.
The arrival time depends on the solar launch time, the velocity history of the ICME front during the transit and also on the shape of the front for ICMEs impacting Earth away from their apex.   This last effect can provide delays up to two days as shown by \citet{Mostl13} with a circular front.   Introducing an ellipsoidal shape could be important away from the apex to better define the front position, but also to better determine the ICME trajectory from imager data (previous paragraph), so finally where the front is crossed.   Then, the ellipsoidal model is expected to improve our forecast abilities as it provides a step forward in modeling the ICME front shape with constraints derived from in situ data.  {A first application of this was performed on a very fast ICME observed by seven spacecraft \citep{Mostl15}. The deduced aspect ratio, $1.4\pm 0.4$, is comparable to the one deduced above from in situ measurements.}

%   {\S\bf Evolution in the heliosphere} \\
  Finally, the present analysis of in situ data can be extended to other solar distances. This requires a consistent set of in situ measurements at different distances away from the Sun. How would the shapes of the MC axis and associated interplanetary shock change with helio-distance? Would the expansion be self-similar for both structures?
In that sense, the future missions Solar Probe Plus and Solar Orbiter will be of great interest to advance our knowledge of the evolution of ICMEs in the inner heliosphere.
   
%%%%%%%%%%%%%%%%%%%%%%%%%%%%%%%%%%%%%%%%%%%%%%%%%%%%%%%%%%%%%%%%%%%%%%%%%%%%%%%
\appendix

\begin{figure*}  %________________________ FIG ______________________________________    
\centering
\IfFileExists{2columnFigures.txt}{
\includegraphics[width=0.48\textwidth,clip]{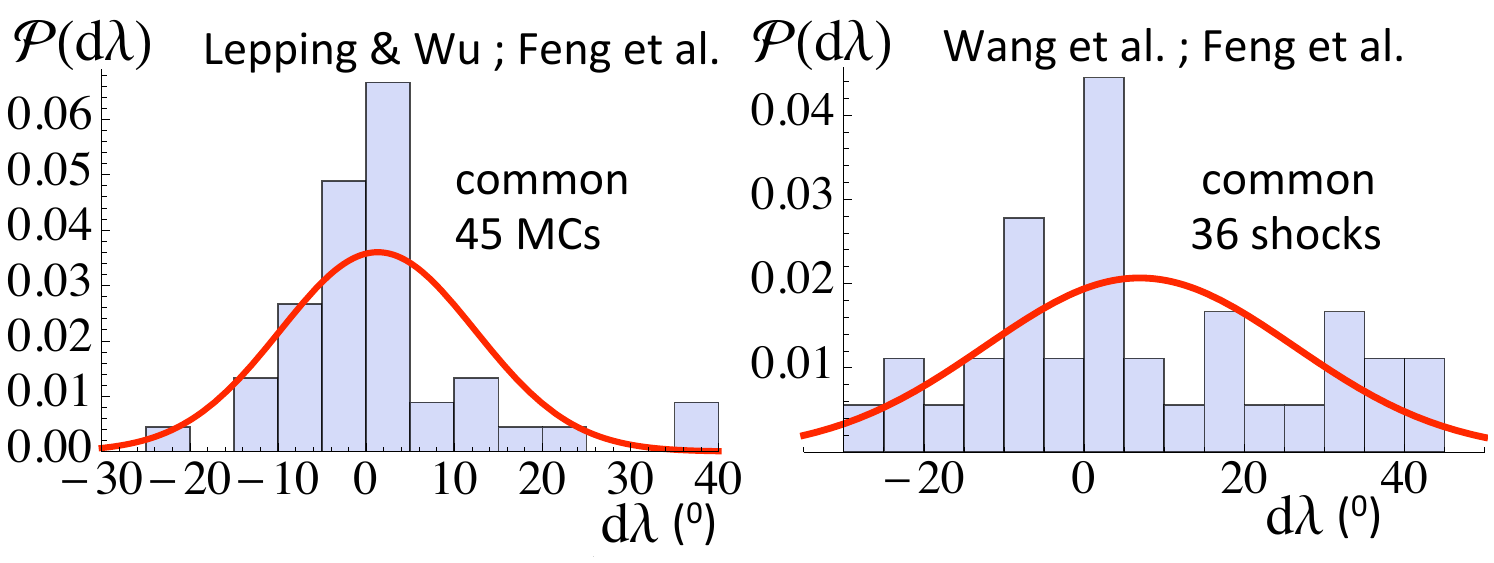}
  } {
\includegraphics[width=0.6\textwidth,clip]{fig_diff_lambda}
}
\caption{Probability distributions of the difference of $\lambda$, {in degree,} obtained by two author-groups studying the same events (left: MCs, right: shocks).  The correlation of the $\lambda$ values are presented in the middle and bottom left panels of Figure~\ref{fig_correlation}.  A Gaussian function with the same mean and standard deviation is superposed in red. \m{These results are used to estimate the error distribution of  $\lambda$ for MCs and shocks. }
}
\label{fig_diff_lambda}
\end{figure*}

\begin{figure*}  %________________________ FIG ______________________________________    
\centering
\IfFileExists{2columnFigures.txt}{
\includegraphics[width=0.8\textwidth,clip]{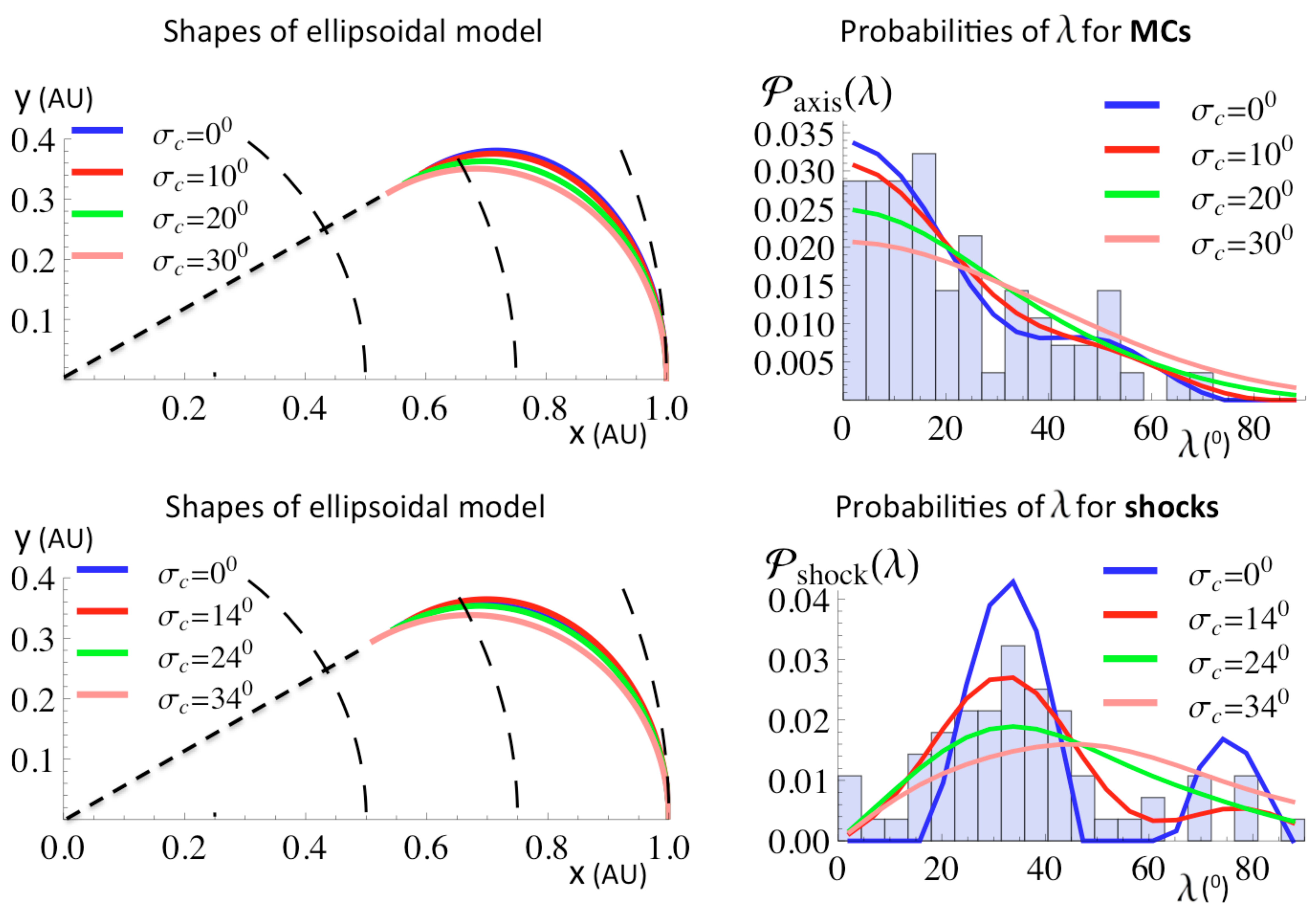}
  } {
\includegraphics[width=0.8\textwidth,clip]{fig_convol_png}
}
\caption{(Half-)Shapes and probability distributions of MC axis (top panels) and shock front (bottom panels) deduced from observations with various dispersions of the MC axis and shock normal.  The observed probability (histogram, Feng \textit{et al.} , 2010) is deconvoluted with a Gaussian kernel, Equation~(\ref{eq_kernel}), with a standard dispersion $\sigma_{\rm d} =10\ensuremath{^\circ}$ for MC and $\sigma_{\rm d} =14\ensuremath{^\circ}$ for shocks (derived from Figure~\ref{fig_diff_lambda}, see Section~\ref{sect_Appendix}). The resulting curve is plotted as a blue curve, rather than an histogram, for superposition.  The other curves are the results of a further convolution by a Gaussian kernel with a standard dispersion $\sigma_{\rm c} $.  The red curves have $\sigma_{\rm c} =\sigma_{\rm d}$ resulting only in a smoothing of the original histogram. The green and pink curves have $\sigma_{\rm c} > \sigma_{\rm d}$ resulting in a broadening of the original histogram.  
\m{The main result is that the deduced MC axis and shock shapes are weakly affected by the errors on $\lambda$ even when these errors are enhanced by a factor 3 compared with the estimations from observations (Figure~\ref{fig_diff_lambda}).}
%{All angles are in degree.}  
}
\label{fig_convol}    
\end{figure*}

\section{
%Implications of errors on the derived axis and shock shapes
{Effects of errors on the derivation of axis and shock shapes}} %%%%%%%%
\label{sect_Appendix}

\subsection{\m{Estimation of the $\lambda$-error distribution}}

% {\S}{\bf --- Quantifies the errors  } \\
The differences between the $\lambda$ values obtained by different authors for the same event were analyzed in Section~\ref{sect_Precision}. 
We quantified the large dispersion of the $\lambda$ values obtained by different authors (Figure~\ref{fig_correlation}), and reported the distribution of the values of the $\lambda$ differences in Figure~\ref{fig_diff_lambda} with histograms. The mean \m{of this  distribution} is negligible for MCs ($\approx 1 \ensuremath{^\circ}$) 
and small for shocks ($\approx 7 \ensuremath{^\circ}$).
 The standard deviation is smaller for MCs ($\sigma_{\rm obs} \approx 11 \ensuremath{^\circ}$) than for shocks ($\sigma_{\rm obs} \approx 19 \ensuremath{^\circ}$). It is remarkable that even with a limited number of common cases (45 MCs and 36 shocks), implying large statistical fluctuations within each histogram bin, both distributions are comparable to a normal (Gaussian) distribution (red curve) with the same mean and standard deviation. 
 
% {\S}{\bf --- Aims: removing the error } \\ 
\m{We analyze below the implications} of $\lambda$ errors on $\mathcal{P}_{\rm obs}(\lambda)$, then on the deduced MC axis and shock shapes.  
To do so, we propose to first estimate the error distributions for MCs and shocks, and to provide a simple model of the error distribution. Then, we attempt to remove these errors on $\mathcal{P}_{\rm obs}(\lambda)$, and finally, we re-introduce variable level of errors in the newly obtained cleaned distribution to study their implications on the deduced MC and shock shapes.

% {\S}{\bf --- Standard deviation of observations } \\ 
The errors in the determination of $\lambda$ are spread in the \m{observed} probability distributions of $\lambda$ (Figure~\ref{fig_prob_lambda}). 
%\m{This spread of the errors therefore makes them less apparent}.  
\m{In order to estimate the $\lambda$ error distribution,
we} decompose $\lambda$ as $\lambda = \lambda_{\rm true} + \lambda_{\rm error}$ where $\lambda_{\rm true}$ is the true $\lambda$ value and $\lambda_{\rm error}$ the error due to the limitations of both observations (data only along the spacecraft trajectory) and modeling (boundary selection and fit of the data to a model, see {Sections \ref{sect_MC-set} and \ref{sect_Shocks-set}).} Considering two samples A and B of common events, we can only access the distribution of $\lambda_{\rm error,A}-\lambda_{\rm error,B}$ (Figure~\ref{fig_diff_lambda}) and not the individual distributions of $\lambda_{\rm error,A}$ and $\lambda_{\rm error,B}$ {(as we do not know $\lambda_{\rm true}$).}  Since the mean values of the $\lambda_{\rm error,A}-\lambda_{\rm error,B}$ distributions are small (Figure~\ref{fig_diff_lambda}), both for MCs and shocks, there are only weak systematic biases.  
This { small} bias can also be observed in the left panels in Figure~\ref{fig_correlation},
with a similar number of cases above or below the identity straight line (brown line){, and also by comparing blue (fitted) and brown lines}.
Next, we assume that $\lambda_{\rm error,A}$ and $\lambda_{\rm error,B}$ are independent statistical variables with the same standard deviation $\sigma_{\rm A} =\sigma_{\rm B}$.
Since the variance of the difference of two independent variables is the sum of their two variances ($\sigma^2= \sigma_{\rm A}^2 +\sigma_{\rm B}^2$), then  $\sigma_{\rm A} =\sigma_{\rm B}$ is a factor $1/\sqrt{2}$ lower than the standard deviation found in Figure~\ref{fig_diff_lambda}.  
For MCs, this implies $\sigma_{\rm A} = \sigma_{\rm B} \approx 8 \ensuremath{^\circ}$, and for shocks $\sigma_{\rm A} = \sigma_{\rm B} \approx 14 \ensuremath{^\circ}$.  
Next, since the distributions of $\lambda_{\rm error,A}-\lambda_{\rm error,B}$ for MCs and shocks are comparable with a normal distribution (Figure~\ref{fig_diff_lambda}), we suppose below that the error distribution of each set of observations is a normal distribution with $\sigma=\sigma_{\rm obs} /\sqrt{2}$:  
 \begin{linenomath}
  \begin{equation} \label{eq_kernel}
  \mathcal{P}_{\rm err}(\lambda) = \frac{1}{\sigma \sqrt{2\pi} }  e^{ - (\lambda /\sigma )^2 / 2 } \, .
  \end{equation}
   \end{linenomath}
In this framework, the observed distribution, $\mathcal{P}_{\rm obs}(\lambda)$, is the result of the true distribution, $\mathcal{P}_{\rm true}(\lambda)$, convoluted with $\mathcal{P}_{\rm err}(\lambda)$ {(called the kernel of the convolution).}
  
\subsection{\m{Deconvolution of the observed probabilities}}

% {\S}{\bf --- Method to deconvolve  } \\ 
The deconvolution of a signal is a delicate problem, in particular since we only have a crude approximation of the kernel with Equation~(\ref{eq_kernel}) \citep[][chap.~13]{numerical-recipes}.  Moreover, there are boundary effects at $\lambda=0 \ensuremath{^\circ}$ and $90 \ensuremath{^\circ}$. In order to limit them we take into account the properties of $\mathcal{P}_{\rm obs}(\lambda)$. At $\lambda=0 \ensuremath{^\circ}$, $\mathcal{P}_{\rm obs}(\lambda)$ is maximum for MCs and close to zero for shocks.  We keep these properties by imposing a symmetric (anti-symmetric) distribution for MCs (shocks), respectively.  In the vicinity of $\lambda=90 \ensuremath{^\circ}$, the distributions are small so we fix $\lambda > 90 \ensuremath{^\circ}$ with zero values on an interval larger than the kernel used. We compare different techniques \m{of deconvolution} present in the Mathematica software (damped least square, Wiener filter and total variation). The first one provides the best results since the estimated $\mathcal{P}_{\rm true}(\lambda)$ distribution has less oscillations and the back convolution of the results by the same kernel is closer to the original distribution.

% {\S}{\bf --- Results of deconvolution then convolution} \\
The results of the deconvolution with $\mathcal{P}_{\rm err}(\lambda)$ with a standard deviation $\sigma_{\rm d}$, for both MCs and shocks studied by \citet{Feng10}, are shown in Figure~\ref{fig_convol} with the blue curves on the right panels.  For MCs, we round $\sigma_{\rm d} \approx 8 \ensuremath{^\circ}$ to $10 \ensuremath{^\circ}$, so closer to the shock value of $\sigma_{\rm d}=14 \ensuremath{^\circ}$. The convolution of these results by the same kernel ($\sigma_{\rm c} =\sigma_{\rm d}$) are shown with the red curves.  They are a smoothed version of the original distributions.  Next, the green and pink curves are the result by a convolution larger by $10 \ensuremath{^\circ}$ and $20 \ensuremath{^\circ}$, respectively ($\sigma_{\rm c}-\sigma_{\rm d}=10 \ensuremath{^\circ}$ and $20 \ensuremath{^\circ}$).   The {distributions $\mathcal{P}(\lambda)$} broaden as $\sigma_{\rm c}$ is increased, as expected. 

\subsection{\m{Implications for the derived MC and shock shapes}}

% {\S}{\bf --- Effect of convolution on the shape  } \\
  We find the best ellipsoidal model for each distribution, similarly to the procedure of Section~\ref{sect_Mod_ellip}.
While the value of $\sigma_{\rm c}$ changes significantly the $\lambda$ distributions, its effect is small on the derived shapes (Figure~\ref{fig_convol}, left panels).   Indeed, there is almost no difference between the shapes derived from $\mathcal{P}_{\rm obs}(\lambda)$ (histograms) and its deconvoluted version (blue curves).  This result could be anticipated from Figure~\ref{fig_ellipsoid} which shows that a large change of $\mathcal{P}(\lambda)$ is needed to have a meaningful change of the derived shape.  Indeed, even further convoluting $\mathcal{P}_{\rm obs}(\lambda)$ with a kernel broader by $10 \ensuremath{^\circ}$ or $20 \ensuremath{^\circ}$, \m{so increasing the error on $\lambda$,} only results in a slightly more bent axis. Quantitatively, for $\sigma_{\rm c}-\sigma_{\rm d}=0 \ensuremath{^\circ}, 10 \ensuremath{^\circ}$ and $20 \ensuremath{^\circ}$ (red, green and pink curves), 
we obtain $b/a=1.3, 1.2$ and $1.1$ for MCs, and $b/a=1.2, 1.1$ and $1.0$ for shocks, respectively.
\m{Comparable results are also found with the cosine model with a weak effect of the $\lambda$ error magnitude on the parameter $n\,f$, then on the deduced shape.}
  
% {\S}{\bf --- effect of deconvolution  } \\
We also explore the effect of a broader kernel for the deconvolution.  Increasing $\sigma_{\rm d}$ implies a deconvoluted $\mathcal{P}(\lambda)$ with more oscillations.  This is a classical result of the deconvolution, with even negative values appearing in a deconvoluted function when the deconvolution kernel is broader than the original function.  This is already present for $\sigma_{\rm d}=14 \ensuremath{^\circ}$ for shocks in Figure~\ref{fig_convol} and the probability $\mathcal{P}(\lambda)$ has been represented with the constraint $\mathcal{P}(\lambda) \geq 0$ and a renormalisation of the probability to get an integral equal unity.  This unphysical negative probability indicates that $\sigma_{\rm d}=14 \ensuremath{^\circ}$ is an over-estimation of the errors (said differently, with such dispersion on the estimations of $\lambda$, $\mathcal{P}_{\rm obs}(\lambda)$ should be broader). \m{With the above results from Figure~\ref{fig_convol}, this further implies that the computed shock shape is weakly affected by errors on $\lambda$.} 
 
%%%%%%%%%%%%%%%%%%%%%%%%%%%%%%%%%%%%%%%%%%%%%%%%%%%%%%%%%%%%%%%%%%%%%%%%%%%%%%%

\begin{acknowledgements}
 {We thank the referees for their careful reading, and whose comments helped improving the manuscript.}
%% Grants
The data used in the present paper are available in \cite{Lynch05}, \cite{Feng10} and \cite{Lepping10} (also available at \url{http://wind.nasa.gov/mfi/mag_cloud_S1.html}) for the magnetic cloud data sets, and in \cite{Feng10} and \cite{Wang10} for the shocks data set. We thank all these authors for making such data publicly available. 
M.J. acknowledges fundings from the Northern Research Partnership with travel support to the Observatoire de Paris.
S.D. acknowledges partial support from the Argentinian grants UBACyT 20020120100220 (UBA), PICT-2013-1462 (FONCyT-ANPCyT), and PIP-11220130100439CO (CONICET). N.~L. was funded by grant NNX13AH94G.
This work was also partially supported by a one-month invitation of P.D. to the Instituto de Astronom\'ia y F\'isica del Espacio, 
and by a one-month invitation of S.D. to the Observatoire de Paris.
%% Conicet
S.D. is member of the Carrera del Investigador Cien\-t\'\i fi\-co, CONICET.
\end{acknowledgements}

 %%% BIBLIOGRAPHY %%%%%%%%%%%%%%%%%%%%%%%%%%%%%%%%%%%%%%%%%%%%%%%%%%%%%%%%%%%

%%%%%%%%%%%%%%%%%%%%%%%%%%%%%%%%%%%%%%%%%%%%%%%%%%%%%%%%%%%%%%%%%%%%%
% format of references provided by the review (.bst)
%\bibliographystyle{aa}
      % file containing the bibtex references (.bib)
%\bibliography{shock}
      % look if the file containing the ``\bibitem'' exits
%\IfFileExists{\jobname.bbl}{}
%{\typeout{}
%\typeout{****************************************************}
%\typeout{****************************************************}
%\typeout{** Please run "bibtex \jobname" to optain}
%\typeout{** the bibliography and then re-run LaTeX}
%\typeout{** twice to fix the references!}
%\typeout{****************************************************}
%\typeout{****************************************************}
%\typeout{}
%}

\end{article}
\end{document}